\documentclass[10pt]{amsart}
\usepackage{graphicx,color}
\usepackage{amsbsy}
\usepackage{amssymb}
\usepackage{mathrsfs}
\usepackage{amsmath}
\usepackage{braket}
\usepackage{cite}

\usepackage{subfig}
\usepackage{soul}

\title{Quantum quenches, sonic horizons and the Hawking radiation in a class of exactly solvable models}

\author{Manuele Tettamanti$^{1}$ \and Sergio L. Cacciatori $^{1,2}$ \and Alberto Parola$^{1}$}

\address{\noindent $^1$Dipartimento di Scienza e Alta Tecnologia, Universit\`a dell' Insubria, via Valleggio 11, 22100 Como, Italy \endgraf
$^2$INFN, Sezione di Milano, I-20133 Milano, Italy}

\begin{document}

\begin{abstract}
Taking advantage of the known exact mapping of the one-dimensional Hard Core Bose (HCB) fluid onto a non-interacting spinless fermion gas, we examine in full detail a thought experiment on cold atoms confined in a quasi-one-dimensional trap, in order to investigate the emergence of the analogue Hawking radiation. The dynamics of a gas of interacting bosons impinging on an external potential is exactly tracked up to the reach of a stationary state. Under few strict conditions on the experimental parameters, the stationary state is shown to be described asymptotically by a thermal distribution, precisely at the expected (analogue) Hawking temperature. However, we find that in most experimental conditions the emerging ``Hawking-like radiation'' is not thermal. This analysis provides a novel many-body microscopic interpretation of the Hawking mechanism, together with useful limits and conditions for the design of future experiments in Bose-Einstein condensates. 
\end{abstract}

\maketitle

\section{Introduction}
General relativity is particularly distinguished for its high internal coherence and, at present days, for its high precision in predictivity, having passed a huge number of tests, which exclude several alternative theories. One of the most recent confirmation is the direct observation of gravitational waves. However, after a century, it still resists to any attempt of conciliation with the principles of quantum physics. An interesting prediction of the combination of gravitational and quantum effects is the result derived by Hawking in 1974, i.e., the fact that black holes are expected to emit a stationary flow of particles at a black-body temperature known as the Hawking temperature, thus evaporating \cite{hawking}. It is a well-known fact, however, that we have no hope of detecting such a phenomenon directly from astrophysical black holes \cite{robert}. Fortunately, Hawking radiation is not related to the dynamics of the gravitational field, but it only depends on the kinematical properties of the spacetime background. In fact, it only relies upon the presence of a horizon and not on whether the black hole space-time metric satisfies Einstein's equations. This fact led Unruh to propose, in 1981, the realization of analogue gravity models \cite{unruh}, in order to reproduce the same kinematical conditions of a black (and/or white) hole in a non-gravitational system. In the years after a plethora of different analogues has been investigated: flowing water, Bose-Einstein condensates (BEC), dilute gases, fiber optics, non-linear dielectrics and others \cite{liberati}. Moreover, several experimental tests have been realized, even though, up to now, a definitive confirmation of the phenomenon is still lacking \cite{sergio}. Among these, experiments with cold atoms BECs look among the most promising ones because of their stability properties and the good level of precision one can reach in the experiments. For both theoretical and technical reasons, these experiments are generally performed in a one-dimensional spatial geometry, corresponding to a two-dimensional spacetime configuration.  In particular, the most celebrated experiments of this kind are those recently performed by Steinhauer \cite{steinhauer1,steinhauer2}, claiming to have detected the Hawking radiation, even if their interpretation is still under debate \cite{numerical,ted1,ted2,ulf}.\\ In these systems, one of the most important difficulties both from a conceptual and a theoretical point of view is due to the fact that each particular analogue model is characterized by non-linear dispersion relations governing the spectrum of the produced quantum modes, differently from the gravitational case. This implies that, in general, it is not possible to compute exactly the spectrum of the emitted quanta (which, of course, is expected to be far from a thermal one) and usually one is led to consider approximate methods in reduced regions of the spectrum. In a Letter \cite{epl}, we recently proposed and summarily discussed an exactly solvable model, devised in order to get exact information on the analogue system and, at the same time, test the limits of precision of the semiclassical methods. Here we present a detailed analysis of the model, proving the results announced in the Letter.

As previously mentioned, the experiments in BECs have always been performed in a quasi-one-dimensional geometry and, therefore, we conform to this limit. We proceed as follows: initially, we consider the analogue of the vacuum of the gravitational case, that is, a stationary flux of interacting bosons. The (repulsive) interaction between particles is essential in order to describe the excitation spectrum in terms of phonons. Accordingly, we model our system as a Tonks-Girardeau gas \cite{girardeau}. Then, we switch on an external potential quenching the system and we let the gas evolve in time. After a while, the system will reach a new stationary state, which, under suitable conditions, may present a sonic horizon, i.e., a point where the flow passes from a subsonic regime to a supersonic one. This evolution stands as the hydrodynamical counterpart of a gravitational collapse that leads to the formation of a horizon. The usual arguments of analogue gravity then predict that, in the final stationary state, a flux of phonons escaping the horizon should be present. Such a flux is expected to be thermal at least in a certain region of the spectrum, with a temperature proportional to the gradient of the differential velocity (the difference between the flow velocity and the sound velocity) at the sonic horizon. These predictions are usually obtained by use of a semiclassical approximation, preventing the complete control of all the physical details playing a role in the phenomenon. Instead, we investigate a microscopic model constituted by interacting quantum particles which is exactly solvable, so that the full dynamics of the model can be followed analytically by use of standard many-body techniques. This is really important in order to understand the limits of validity of the semiclassical treatment on which analogue gravity arguments are based. To his purpose, we reproduce analytically an ideal experiment step by step, so to make evident the emergence of the Hawking radiation. We also compare the exact results with the ones obtained by applying the semiclassical approximation to our model and we discuss possible experimental set-ups able to provide unambiguous evidence of the presence of the analogue Hawing radiation in Bose-Einstein condensates. 

A closely related model was proposed some years ago \cite{giovanazzi} in the analogue gravity framework and, 
by use of semiclassical arguments, the occurrence of thermal Hawking radiation in the asymptotic state was correctly predicted. Here, through the exact solution of the model, we can follow the full dynamics of the system beyond the semiclassical approximation, providing quantitative estimates for the relevant physical observables.

\section{Hard Core Bose gas}
\subsection{Bose-Fermi mapping}
Exactly solvable models of interacting bosons in one dimension have been known for many years \cite{mattis}. The most celebrated off-lattice system in this class is the $\delta$-interacting Bose gas \cite{lieb} defined by the Hamiltonian
\begin{eqnarray}
\hat H =\int dx &\Big [ \frac{\hbar^2}{2m}\,\partial_x\hat\psi^\dagger(x)  \partial_x \hat\psi(x) + 
+\frac{1}{2}\,g\,\hat\psi^\dagger(x)\hat\psi(x)
\hat\psi^\dagger(x)\hat\psi(x)\Big ] \, ,
\label{h}
\end{eqnarray}
where $\hat\psi(x)$ is the usual bosonic annihilation field operator \cite{fetter}. Although the full spectrum can be calculated via Bethe-Ansatz for arbitrary interaction strength $g$ \cite{lieb,lieb2}, the structure of the wavefunction is very cumbersome and few analytical results can be effectively extracted. Studying the dynamics of this system in an external potential is even more challenging and only recently some progress has been made in this direction \cite{andrei1,andrei2}. However, in the particular (but physically relevant) limit of hard core repulsion between bosons (i.e., $g\to\infty$ in Eq. (\ref{h})) the problem becomes remarkably simple and the interacting Hamiltonian can be exactly mapped into that of a free Fermi gas \cite{girardeau}, making the analysis considerably easier. The Bose-Fermi correspondence, valid only in one dimension, is based on two observations: $i)$ choosing a basis set labeled by all possible particle configurations in real space, the matrix elements of the kinetic term of the Hamiltonian (\ref{h}) do not depend whether the field operators $\hat\psi(x)$ commute or anti-commute; $ii)$ the hard core constraint is automatically satisfied by fermions. Due to this mapping, which holds also  in the presence of arbitrary external potentials, the full energy spectrum and all the static and dynamic correlation functions involving only density operators of the Bose gas coincide with those of the free fermion model \cite{note}. Therefore, we are led to conclude that the Hard Core Bose fluid (HCB), also known as Tonks-Girardeau gas, represents the minimal model including the essential features of the interacting Bose gas and allowing for an exact analytical and manageable solution. It is also worth noting that, although the hard core interaction is certainly a limiting case of the more general Hamiltonian (\ref{h}), the physical properties of the interacting Bose gas change smoothly up to $g\to \infty$, which does not correspond to a singular limit \cite{lieb}. Indeed, recent studies on the quench dynamics of the interacting Bose gas showed that the HCB limit faithfully represents the generic behavior of the model \cite{andrei1,andrei2}. Most importantly, the HCB limit can be reached in suitably designed experiments with cold atoms (namely $^{87}$Rb) as proved in Refs. \cite{kinoshita,bloch}. The properties of this model are here investigated analytically and numerically at vanishing physical temperature, in order to unambiguously identify the thermal nature of the emerging Hawking radiation.
\subsection{Quantum averages}
Before exploring the physical content of the model, we first recall the explicit expressions useful to evaluate the most relevant properties we are going to investigate. Each eigenstate of a single free fermion in an external potential (which does not support bound states) is uniquely labeled by a ``wavevector'' $k$ characterizing the asymptotic scattering state \cite{landau}. The exact eigenstates of a collection of non-interacting fermions are then written as Slater determinants of single particle wavefunctions and are labeled by the momentum distribution $f(k)$, defining the set of occupied states. In particular, the ground state is obtained by filling all the energy levels up to the Fermi wavevector $|k| < k_F$. The quantum averages of one- and two-body operators \cite{fetter} for a given arbitrary momentum distribution function $f(k)$ are defined in the following way: 
\par\noindent 
$\bullet \,$ the average of any one-body operator $O_1=\sum_{n=1}^{N} O_1(x_n,-i\hbar\partial_{x_n})$ is written in terms of the single particle eigenstates $|k\rangle$ as:
\begin{equation}
\langle O_1 \rangle = \int_{-\infty}^\infty dk\, \langle k| O_1(x,-i\hbar\partial_x)|k\rangle \, f(k)\,.
\label{1b}
\end{equation}
\par\noindent 
$\bullet \,$ For a two-body operator $O_2$ expressed in terms of density operators:
\begin{eqnarray}
\label{2b}
\langle O_2 \rangle &&= \int_{-\infty}^\infty dk\,dk^\prime\,
\Big [ \langle k,k^\prime|O_2(x;x^\prime)+ |k,k^\prime\rangle
-\langle k,k^\prime|O_2(x;x^\prime)|k^\prime,k\rangle \Big ] \, f(k)\,f(k^\prime)\,.
\end{eqnarray}

\section{Quench dynamics}
Following the line of thought originally devised by Hawking in the astrophysical context \cite{hawking}, we discuss a ``Gedankenexperiment'' in order to detect the analogue Hawking radiation in Bose-Einstein condensates. First, let us consider a single free quantum particle initially set in a plane-wave stationary state of wavevector $k$
\begin{equation}
\psi^0_k(x) = \frac{e^{ikx}}{\sqrt{2\pi}} \, .
\end{equation}
Then, at $t=0$, an external repulsive potential $V(x)$ is suddenly switched on. The ensuing time evolution is formally written as 
\begin{eqnarray}
\label{psit}
\psi_k(x,t) &=& \int_{-\infty}^\infty dp\, \langle \phi_p| \psi^0_k \rangle \,\phi_p(x)\,e^{-\frac{i}{\hbar}\epsilon_pt}, \\
\langle \phi_p| \psi^0_k \rangle &=& \int_{-\infty}^{\infty} \frac{dx}{\sqrt{2\pi}} \, \phi^*_p(x)\,e^{ikx}\,e^{-\eta |x|},
\label{internal}
\end{eqnarray}
where $\phi_p(x)$ are the exact eigenfunctions of a particle in the external potential $V(x)$, $\epsilon_p=\frac{\hbar^2p^2}{2m}$ are the associated eigenvalues and $\eta\to 0^+$ is the usual convergence factor. Now, let us look at the long time evolution at fixed position $x$. A careful analysis of these expressions shows that, for each $k$, $\psi_k(x,t)$ is given by the sum of an exact scattering eigenfunction of the system in the presence of the external potential multiplied by a time dependent phase factor, plus a contribution which vanishes as $t\to+\infty$. These two terms represent, respectively, the asymptotic stationary state and a traveling wave originated during the quench (see also Appendix \ref{app:longtime}). 

If we now take a HCB gas instead of a single particle, according to the Bose-Fermi mapping we can equivalently consider a free spinless Fermi gas, whose dynamics is described by a Slater determinant of single-particle wavefunctions. Therefore, to describe the quench dynamics in a HCB gas, we start from a Fermi gas filling all the single-particle states up to the Fermi momentum $k_F$. Then we set the gas into motion to the left by rigidly shifting the momentum distribution which now includes the states $-k_F-k_0 < k < k_F-k_0$. In the following we will always consider $k_0 < k_F$, which corresponds to a subsonic flow. Note that uniform density $\rho_0$ and the initial velocity $v_0$ of the fluid are simply related to $k_F$ and $k_0$ by $\rho_0 = \frac{k_F}{\pi}$ and $v_0 = -\frac{\hbar k_0}{m}$, while the corresponding sound velocity is $c= \frac{\hbar k_F}{m}$. Lastly, we turn on the external potential and wait, at fixed position $x$, until a stationary state is reached. During the unitary time evolution, the many-body wavefunction preserves its Slater determinant structure with time-dependent orbitals given by (\ref{psit}). To quantitatively discuss the long time properties of the system we have to specify the form of the external potential $V(x)$. In the following, we consider two representative model potentials often investigated both theoretically and experimentally.  

\subsection{Step potential}
We first consider an external potential of the form $V(x)=V_0\Theta(x)$ where $\Theta$ is the Heaviside function and we write for convenience $V_0=\frac{\hbar^2Q^2}{2m}$. This ``waterfall'' potential \cite{nota-step} coincides with the original proposal by Unruh and has also been adopted in the experiments performed by Steinhauer \cite{steinhauer1,steinhauer2}. 

A set of single-particle eigenfunctions are explicitly reported in Appendix \ref{app:spectrum}. With this choice, by evaluating the long time limit of Eq. (\ref{psit}), the stationary state is defined by the single-particle orbitals: 
\begin{equation}
\psi_k(x,t) \to 
\begin{cases}
\phi_k(x)\,e^{-i\frac{\hbar k^2}{2m}t} & 
k>0 \\
\sqrt{\frac{|k|}{p}}\,\phi_{-p}(x)\,
e^{-i\frac{\hbar p^2}{2m}t} & 
k<0 
\end{cases} \, ,
\label{asy}
\end{equation}
with $p=\sqrt{k^2+Q^2}$. A sketch of the derivation of this asymptotic expression is reported in Appendix \ref{app:longtime}. The momentum distribution defining the Slater determinant remains unchanged during the evolution and, therefore, the stationary state is defined by the orbitals belonging to the interval $-k_F-k_0<k<k_F-k_0$. As a consequence, after an initial transient, an observer at a given position $x$ will perceive the moving fluid in a stationary state described by the asymptotic single-particle wavefunctions (\ref{asy}). Note that, while the full evolution (\ref{psit}) must conserve both the energy and the total number of particles, the stationary state orbitals do not have to, since the waves formed during the quench and traveling in both directions may carry energy (and particles) to infinity at long times. Moreover, in the non-uniform stationary state, the local energy and momentum density may differ even in the asymptotic regions $x\to\pm\infty$ from their initial value before the quench. 

\begin{figure}
\begin{center}
\includegraphics[width=7cm]{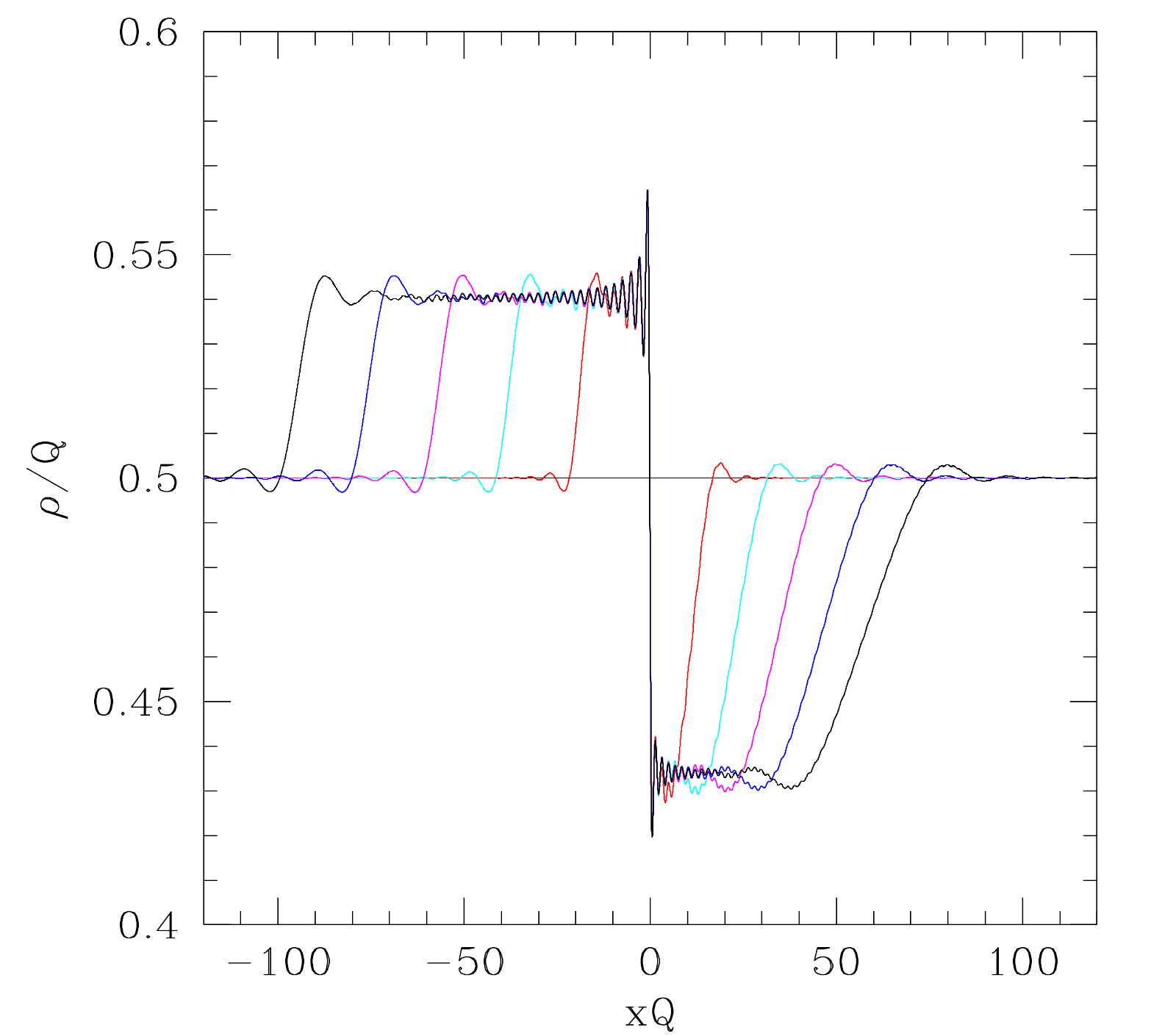}
\caption{Density profile of the HCB gas after the quench for $k_F=\frac{\pi}{2}\,Q$ and $k_0=\frac{3\pi}{50}\,Q$.
Initially the fluid has a uniform density $\rho_0=0.5\,Q$. Colors refer to different times after the quench: red, cyan, magenta, blue, black. The time lapse between curves is $10\,\frac{m}{\hbar Q^2}$.}
\label{fig-evol-1402}
\end{center}
\end{figure}

\begin{figure}[ht!] 
\begin{center}
\includegraphics[width=7cm]{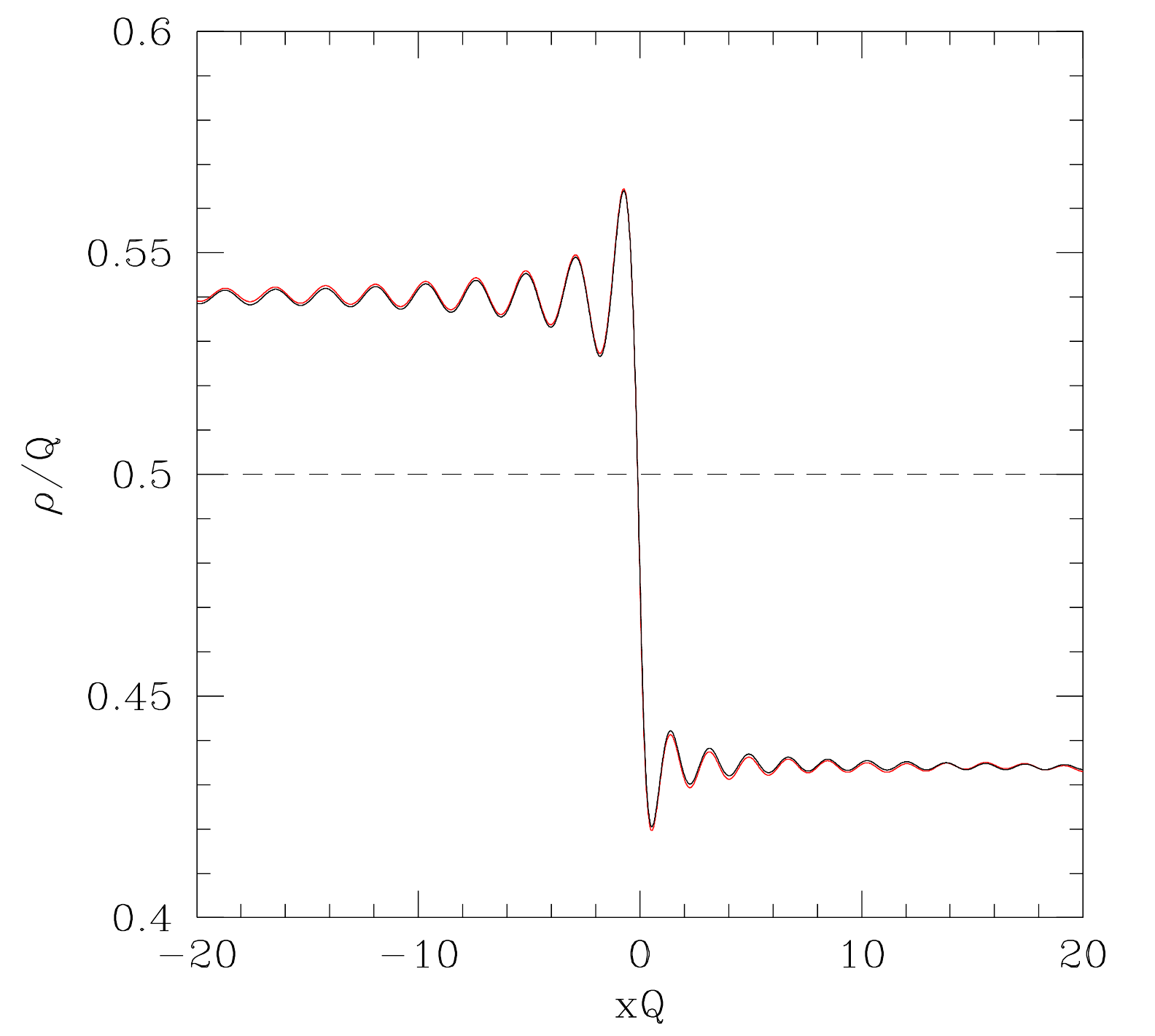}
\caption{Blow up of the long time results shown in Fig. \ref{fig-evol-1402}. Density profile of the HCB gas long after the quench. The analytical stationary solution (black line) is compared to the numerical solution of the Schr\"odinger equation after a time $50\,\frac{m}{\hbar Q^2}$ (red curve); in the figure, the two curves are superimposed. The dashed line shows the value of the density before the quench.}
\label{fig-rho-1402}
\end{center}
\end{figure}

The stationary state properties can now be analytically evaluated by use of Eq. (\ref{asy}). In Fig. \ref{fig-evol-1402} we show a few snapshots of the time evolution of the density profile 
\begin{equation}
\rho(x,t) = \int_{-k_F-k_0}^{k_F-k_0} dk\, \vert \psi_k(x,t)\vert^2 \, ,
\label{dens}
\end{equation}
as obtained by numerical integration of the Schr\"odinger equation for the Fermi gas flowing in a waterfall potential.  
Lengths are expressed in units of $Q^{-1}$ and times in units of $\tau=\frac{m}{\hbar Q^2}$. Here the initial density is set to $\rho_0=0.5\, Q$ and the initial velocity is $0.12$ times the sound speed of the fluid $c_0= \frac{\hbar}{m}\pi\rho_0$.  The quench dynamics is clearly seen in the figure: two waves are generated at $t=0$ in $x=0$ and propagate at different velocities $ c_0\pm \vert v_0 \vert =\frac{\hbar}{m}(k_F\pm k_0)$ in the downstream ($+$) and upstream ($-$) direction. In the central region the density profile shows a rapid variation near the waterfall and several undulations develop, the density being lower upstream than downstream. These waves cannot be interpreted in terms of the usual dispersive shock waves (DSW) or simple waves, as they disappear if the same dynamics is studied by means of the mean-field Gross-Pitaevskii equation. Furthermore, as Fig. \ref{fig-rho-1402} shows, they persist in the stationary state, differently from what we would expect in the case of DSW. They thus reflect the presence of the hard core constraint. In Fig. \ref{fig-rho-1402}, a blow-up of the density profile in the inner region at long times $t=50\,\tau$ is compared to the analytical solution based on Eq. (\ref{asy}). As it can be seen, the agreement is remarkable.

\begin{figure} [ht!]
\begin{center}
\includegraphics[width=7cm]{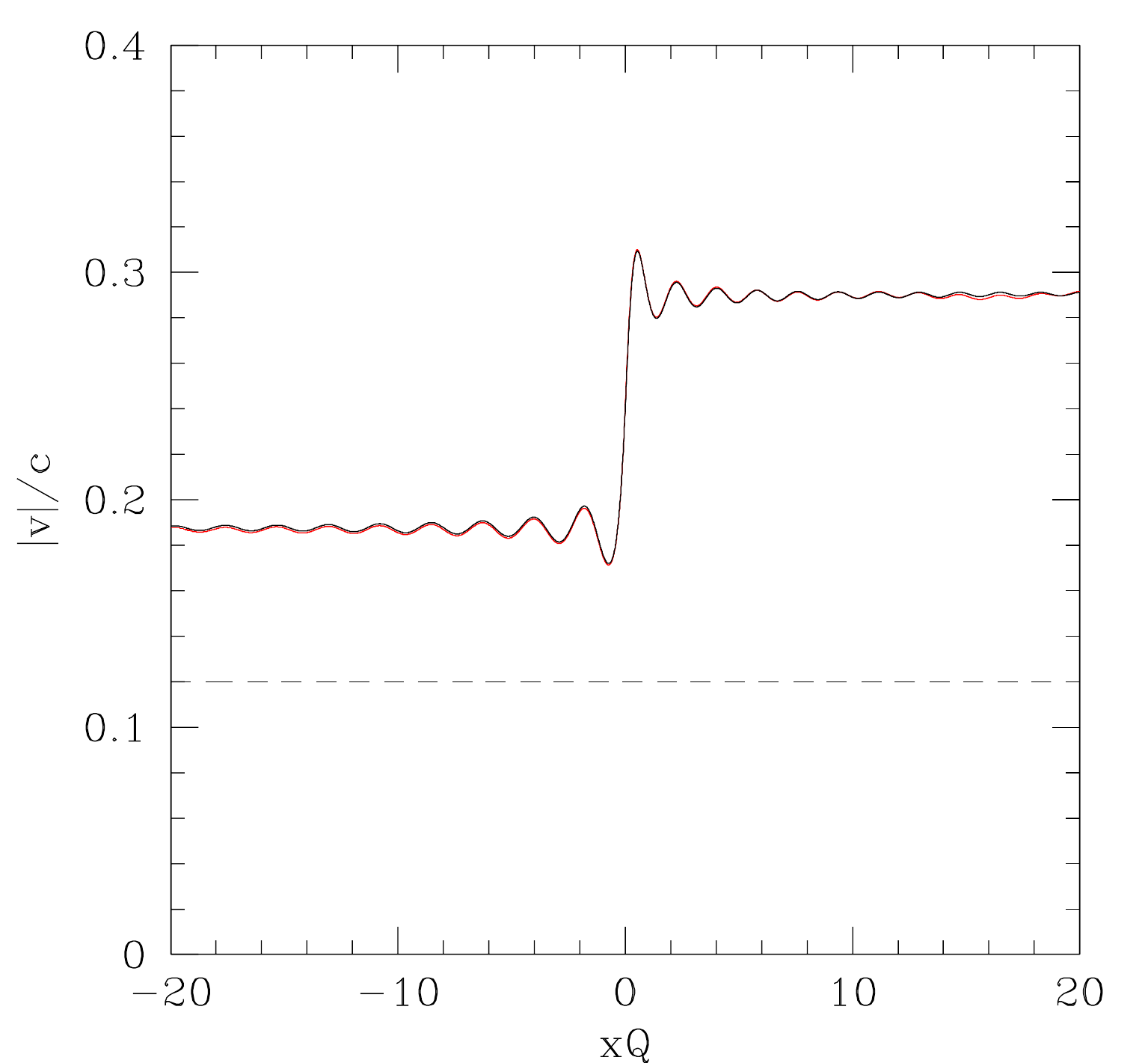}
\caption{Ratio between the absolute value of the fluid velocity and the local sound speed long after the quench for the same parameters of Fig. \ref{fig-evol-1402}. The dashed line shows the ratio before the quench. The analytical stationary solution (black line) is compared to the numerical solution of the Schr\"odinger equation after a time $50\,\frac{m}{\hbar Q^2}$ (red curve); in the figure, the two curves are superimposed.}
\label{fig-vel-1402}
\end{center}
\end{figure}

The velocity profile of the Fermi gas is defined in terms of the local mass flux 
\begin{equation}
j(x,t)  = \Re\,\left [ i\,\hbar\,
\int_{-k_F-k_0}^{k_F-k_0} dk\, \psi_k(x,t)\,\partial_x\psi_k^*(x,t)\right ] 
\nonumber
\end{equation}
as $v(x,t) = \frac{j(x,t)}{m\rho(x,t)}$. In Fig. \ref{fig-vel-1402}, the fluid velocity is compared to the local sound speed, which, for a HCB gas, is given by $c(x,t) = \frac{\hbar}{m}\pi\rho(x,t)$. 
For this parameter choice the flow is always subsonic, the ratio being less than unity. The excellent agreement between the numerical results and the analytical expressions confirms the correctness of the theoretical analysis.

\begin{figure} [ht!]
\begin{center}
\includegraphics[width=7cm]{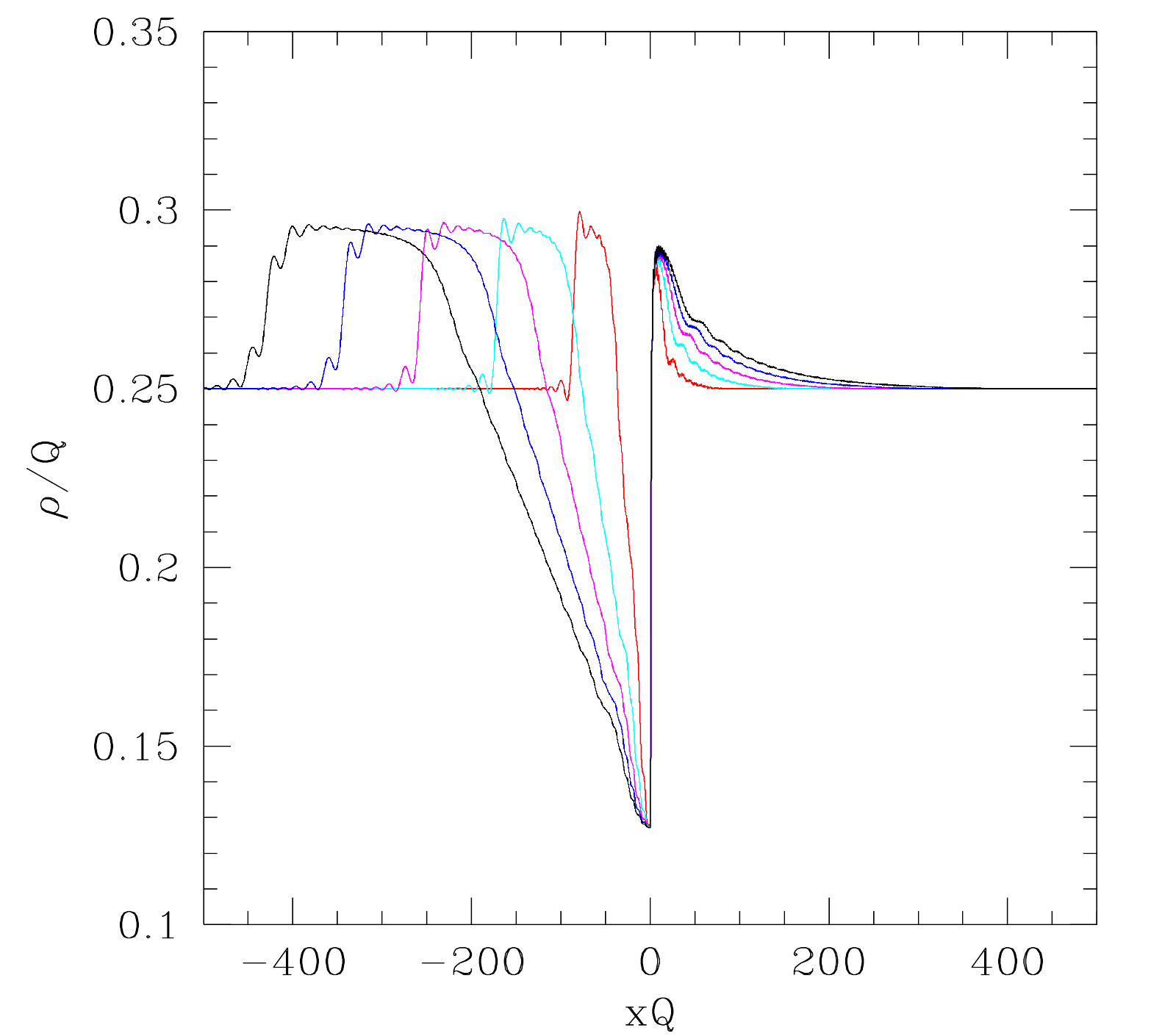}
\caption{Density profile of the HCB gas after the quench for $k_F=k_0=\frac{\pi}{4}\,Q$. Initially the fluid has a uniform density $\rho_0=0.25\, Q$. Colors refer to different times after the quench: red, cyan, magenta, blue, black. The time lapse between curves is $50\,\frac{m}{\hbar Q^2}$.}
\label{fig-evol-0808}
\end{center}
\end{figure}

\begin{figure} [ht!]
\begin{center}
\includegraphics[width=7cm]{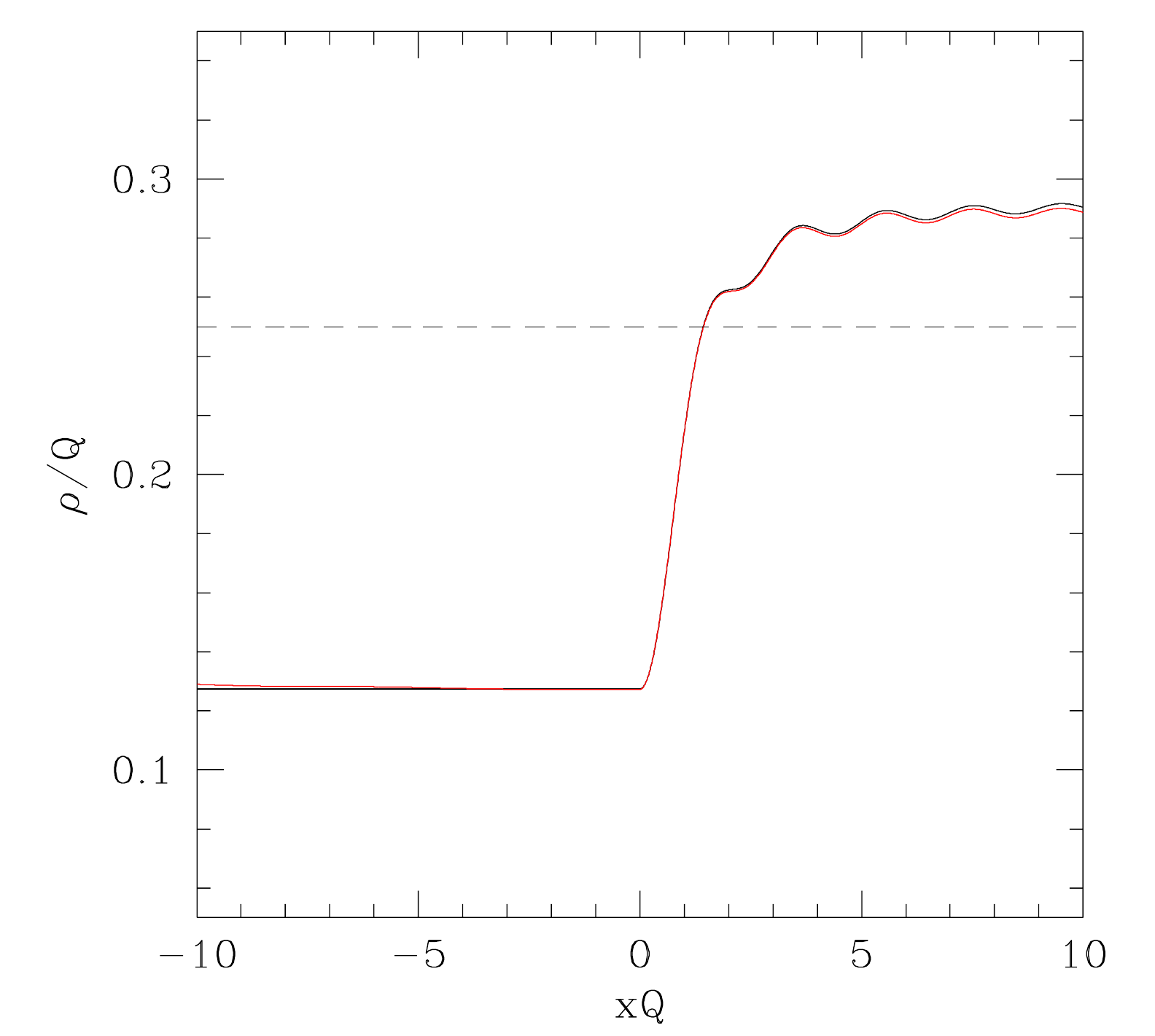}
\caption{Blow up of the long time results shown in Fig. \ref{fig-evol-0808}.
Density profile of the HCB gas long after the quench. The analytical stationary solution (black line) is compared to the numerical solution of the Schr\"odinger equation after a time $250\,\frac{m}{\hbar Q^2}$ (red curve). The dashed line shows the value before the quench.}
\label{fig-rho-0808}
\end{center}
\end{figure}

\begin{figure} [ht!]
\begin{center}
\includegraphics[width=7cm]{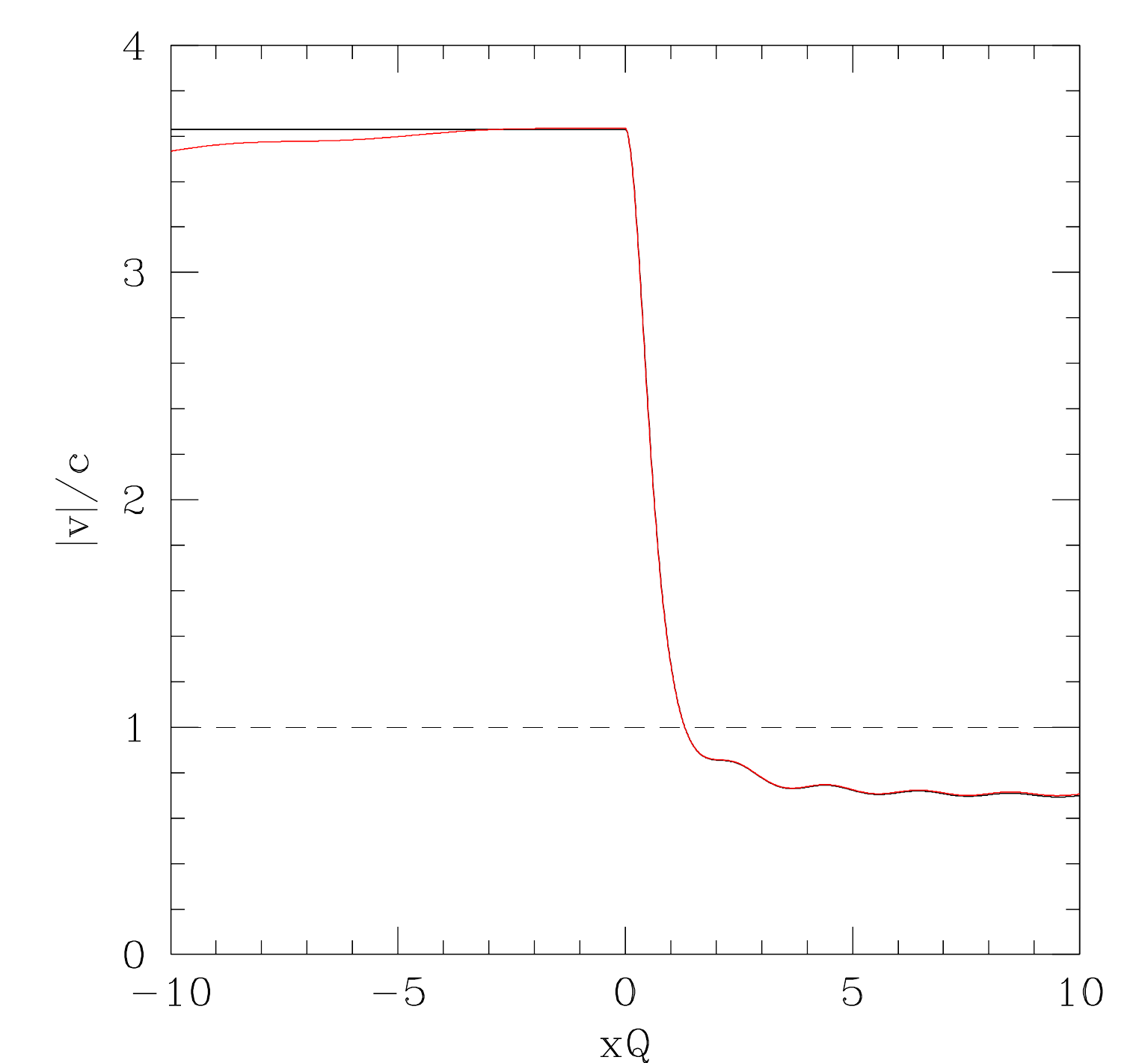}
\caption{Ratio between the absolute value of the fluid velocity and the local sound speed long after the quench for the same parameters of Fig. \ref{fig-evol-0808}. The dashed line shows the ratio before the quench. The analytical stationary solution (black line) is compared to the numerical solution of the Schr\"odinger equation after a time $250\,\frac{m}{\hbar Q^2}$ (red curve).}
\label{fig-vel-0808}
\end{center}
\end{figure}

At fixed initial density $\rho_0$, for sufficiently small values of the initial velocity $v_0$ the stationary state is subsonic while, for larger values of $v_0$, a supersonic region appears near the potential step, although the flow remains subsonic in the asymptotic region $x\to -\infty$. Nevertheless, a sonic horizon is present whenever $k_F-k_0 <Q$. Further increasing $v_0$ the flow becomes fully supersonic beyond the sonic horizon located close to $x=0$. 
\begin{figure} [ht!]
\begin{center}
\includegraphics[width=7cm]{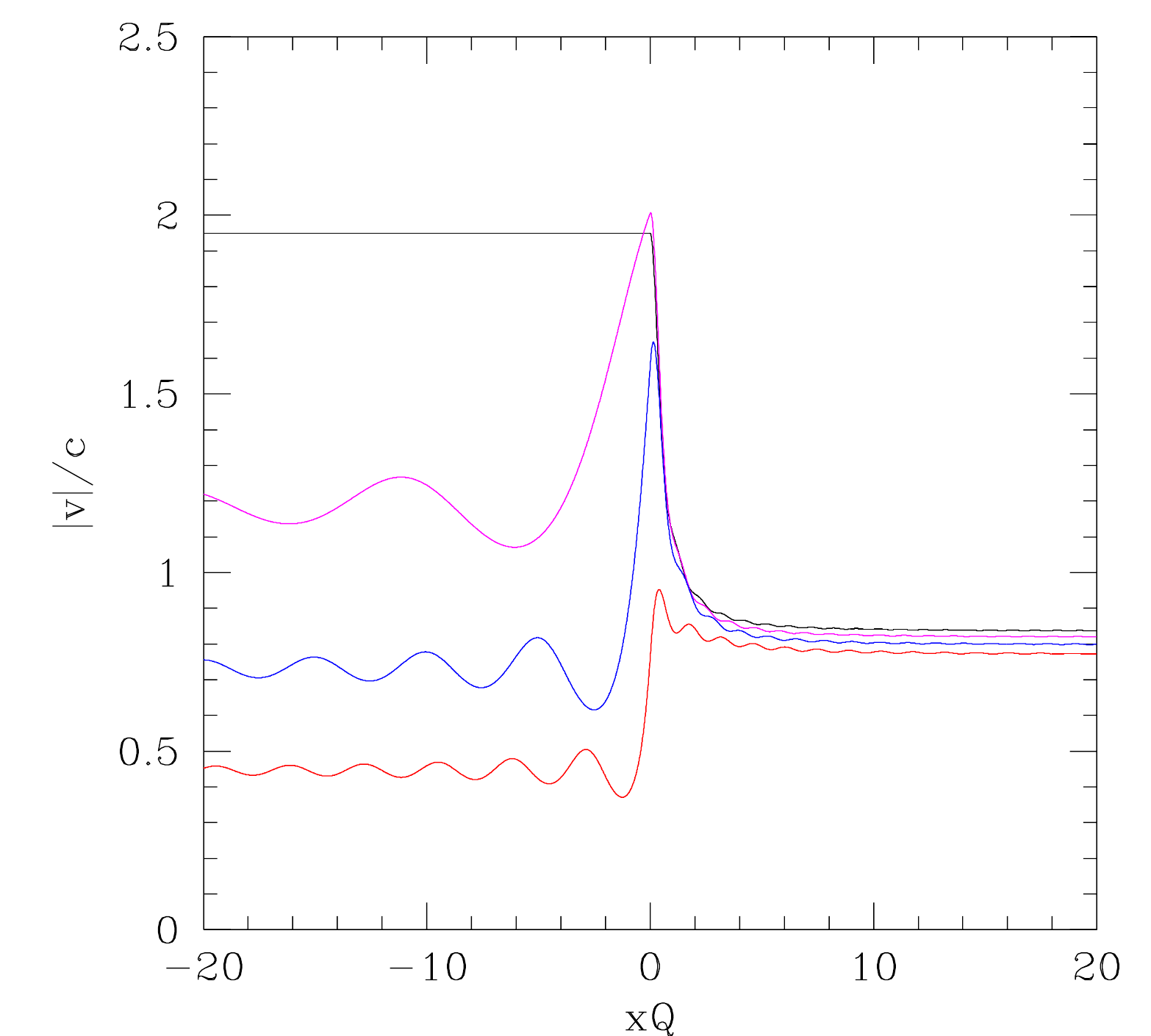}
\caption{Ratio between the absolute value of the fluid velocity and the local sound speed in the stationary state reached starting from $\rho_0=0.5\, Q$ and initial velocities $v_0=-\frac{5\pi}{10} \frac{\hbar Q}{m}$ (black curve), $v_0=-\frac{4\pi}{10} \frac{\hbar Q}{m}$ (magenta), $v_0=-\frac{3\pi}{10} \frac{\hbar Q}{m}$ (blue), $v_0=-\frac{2\pi}{10} \frac{\hbar Q}{m}$ (red). The plot shows that, for a fixed value of the initial density $\rho_0$, by increasing the initial velocity we move from a totally subsonic configuration in the $x<0$ region (red curve), to configurations which are totally supersonic (magenta and black curves).}
\label{fig-vel-tot}
\end{center}
\end{figure}
Fig. \ref{fig-vel-tot} shows the ratio between the fluid velocity and the local sound speed in the stationary state for few values of the initial parameters $(k_F,k_0)$. The behavior is generally non monotonic and characterized by undulations in the downstream region, except for $k_0=k_F$ (i.e., for $v_0=-\pi\frac{\hbar\rho_0}{m}$) when the oscillations disappear and the velocity becomes constant beyond the horizon. 
A few snapshots of the HCB dynamics are shown in Fig. \ref{fig-evol-0808} for a set of parameters triggering a supersonic transition.
The analytical stationary state and numerical results long after the quench are compared in Fig. \ref{fig-rho-0808} and the agreement is remarkable. Note that, in this case, the development of the stationary state requires considerable longer times and the numerical integration was carried out up to times as large as $t=250\,\tau$ in order to obtain a stationary solution in the range $|x| \lesssim 10\, Q^{-1}$.
The Mach ratio $|v|/c$ long after the quench is shown in Fig. \ref{fig-vel-0808} displaying a supersonic transition. For this choice of parameters at $t=0$ the fluid velocity coincides with the sound speed of the uniform gas. 

\subsection{Repulsive barrier}
We now turn to a different form of the external potential: a repulsive barrier given by
\begin{equation}
V(x) = \frac{V_0}{\cosh(\alpha x)^2} \, ,
\label{cosh}
\end{equation}
where the parameter $\alpha$ governs the width and smoothness of the barrier, while $V_0$ is conveniently parametrized as
$V_0=\frac{\hbar^2 Q^2}{2m}+\frac{\hbar^2 \alpha^2}{8m}$ \cite{error}. Taking advantage of the exact solution of the eigenvalue problem \cite{landau} a set of exact, properly normalized eigenfunctions can be chosen as: 
\begin{eqnarray}
\phi_k(x) &= \frac{\Gamma(\frac{1}{2}-i\frac{k+Q}{\alpha})\Gamma(\frac{1}{2}-i\frac{k-Q}{\alpha})}
{\sqrt{2\pi}\,\Gamma(1-i\frac{k}{\alpha})\Gamma(-i\frac{k}{\alpha})}
\left [\zeta(1-\zeta)\right]^{-i\frac{k}{2\alpha}} F\left (\frac{1}{2}-i\frac{k+Q}{\alpha}, \frac{1}{2}-i\frac{k-Q}{\alpha};1-i\frac{k}{\alpha}; \zeta\right ), \quad 
\label{hyper}
\end{eqnarray}
where $\zeta=\frac{1-\tanh(\alpha x)}{2}$ and $\Gamma(a)$, $F(a,b;c;\zeta)$ are the usual Gamma and Hypergeometric functions, respectively \cite{abramo}. The above expression, valid for $k>0$, represents a right-moving scattering solution with energy $\epsilon_k = \frac{\hbar^2k^2}{2m}$. The degenerate eigenfunction with $k<0$ is obtained by replacing $k\to -k$ and $x\to -x$. We can proceed as previously discussed by setting the free Fermi gas in a Slater determinant of plane waves with momenta in the interval $-k_F-k_0 < k < k_F-k_0$ (with $0 < k_0 < k_F$). An analysis similar to the one carried out for the step potential (see also Appendix \ref{app:longtime}) shows that, with this choice of eigenfunctions, by switching on the barrier and waiting for equilibration, the system relaxes to a state defined by a Slater determinant of the eigenstates (\ref{hyper}) with momenta belonging to the same interval. This set of eigenstates, therefore, describes the asymptotic stationary state of the system long after the quench. The analytical form of the asymptotic densities at $x\to\pm\infty$ are reported in Appendix \ref{app:exact}. 

\begin{figure} [ht!]
\begin{center}
\includegraphics[width=7cm]{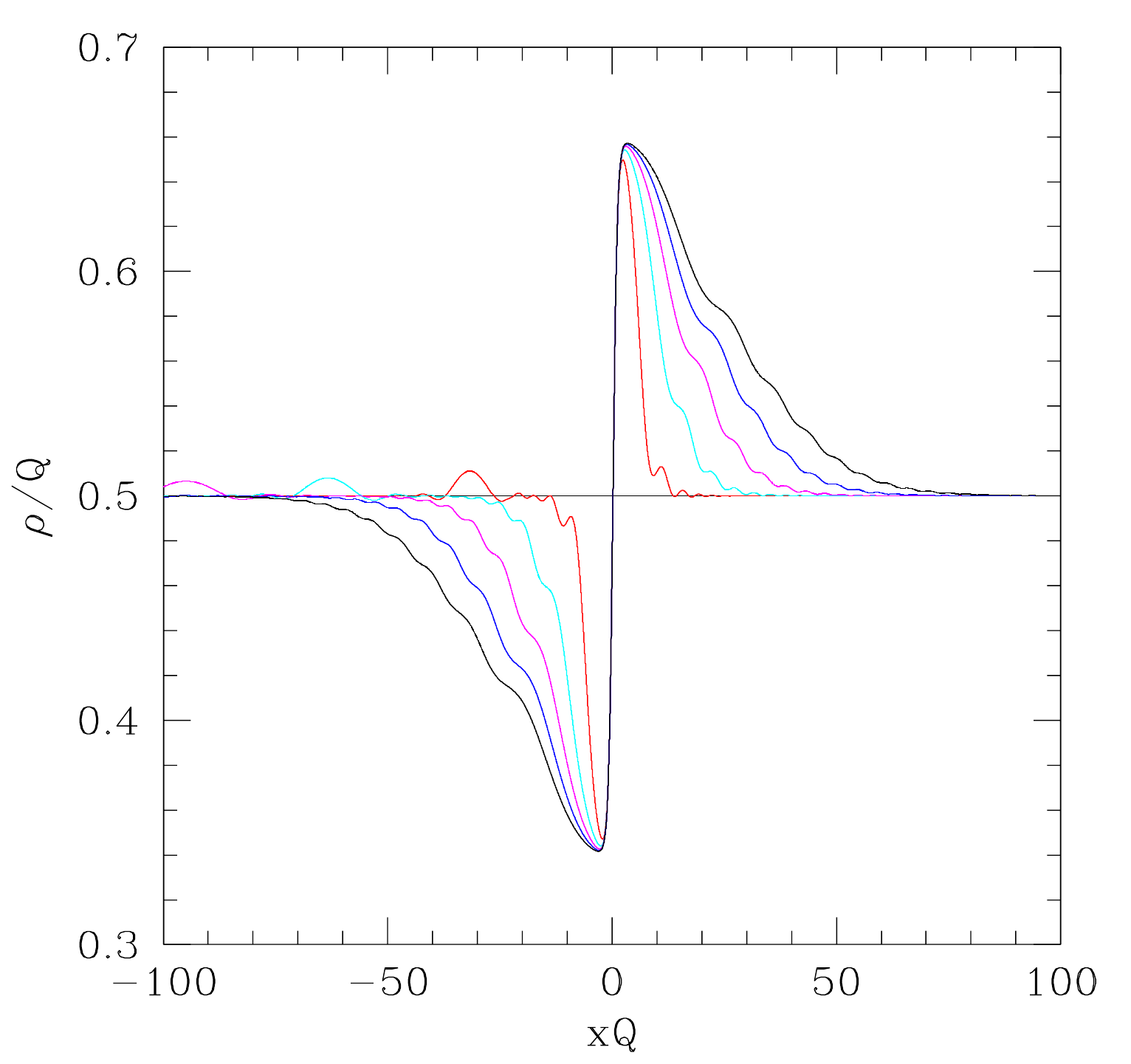}
\caption{Density profile of the HCB gas after the quench for $k_F=k_0=\frac{\pi}{2}\,Q$ and $\alpha=Q$. Initially the fluid has a uniform density $\rho_0=0.5\, Q$. Colors refer to different times after the quench: red, cyan, magenta, blue, black. The time lapse between curves is $10\,\frac{m}{\hbar Q^2}$.}
\label{fig-evol-a1}
\end{center}
\end{figure}

\begin{figure} [ht!]
\begin{center}
\includegraphics[width=7cm]{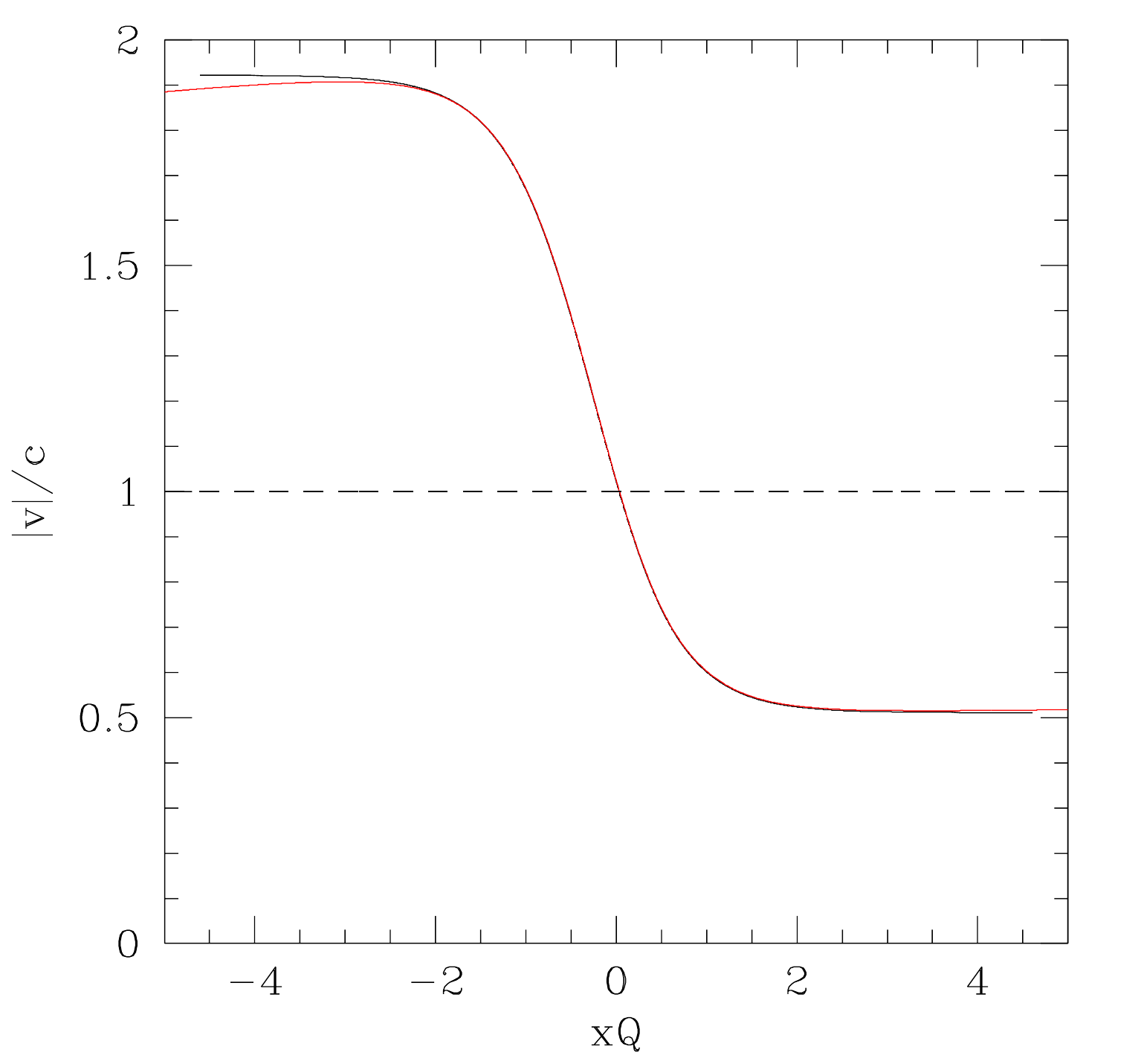}
\caption{Ratio between the absolute value of the fluid velocity and the local sound speed for the same parameter choice as Fig. \ref{fig-evol-a1}. The dashed line shows the ratio before the quench. The analytical stationary solution (black line) is compared to the numerical solution of the Schr\"odinger equation after a time $50\,\frac{m}{\hbar Q^2}$ (red curve).}
\label{fig-evol-a2}
\end{center}
\end{figure}

\begin{figure} [ht!]
\begin{center}
\includegraphics[width=7cm]{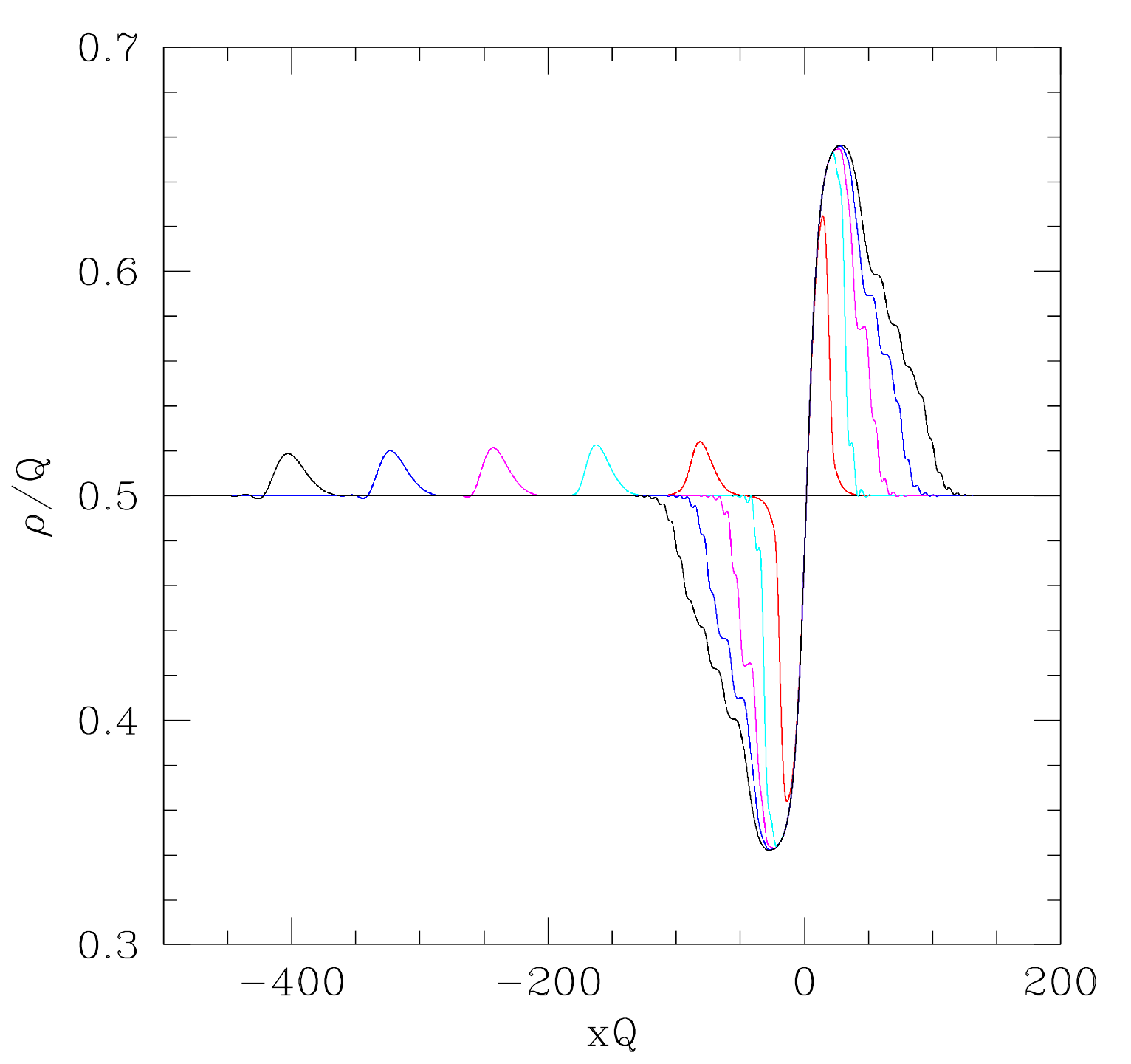}
\caption{Density profile of the HCB gas after the quench for $k_F=k_0=\frac{\pi}{2}\,Q$ and $\alpha=0.1\,Q$. Initially the fluid has a uniform density $\rho_0=0.5\, Q$. Colors refer to different times after the quench: red, cyan, magenta, blue, black. The time lapse between curves is $25\,\frac{m}{\hbar Q^2}$.}
\label{fig-evol-a3}
\end{center}
\end{figure}

\begin{figure}[ht!] 
\begin{center}
\includegraphics[width=7cm]{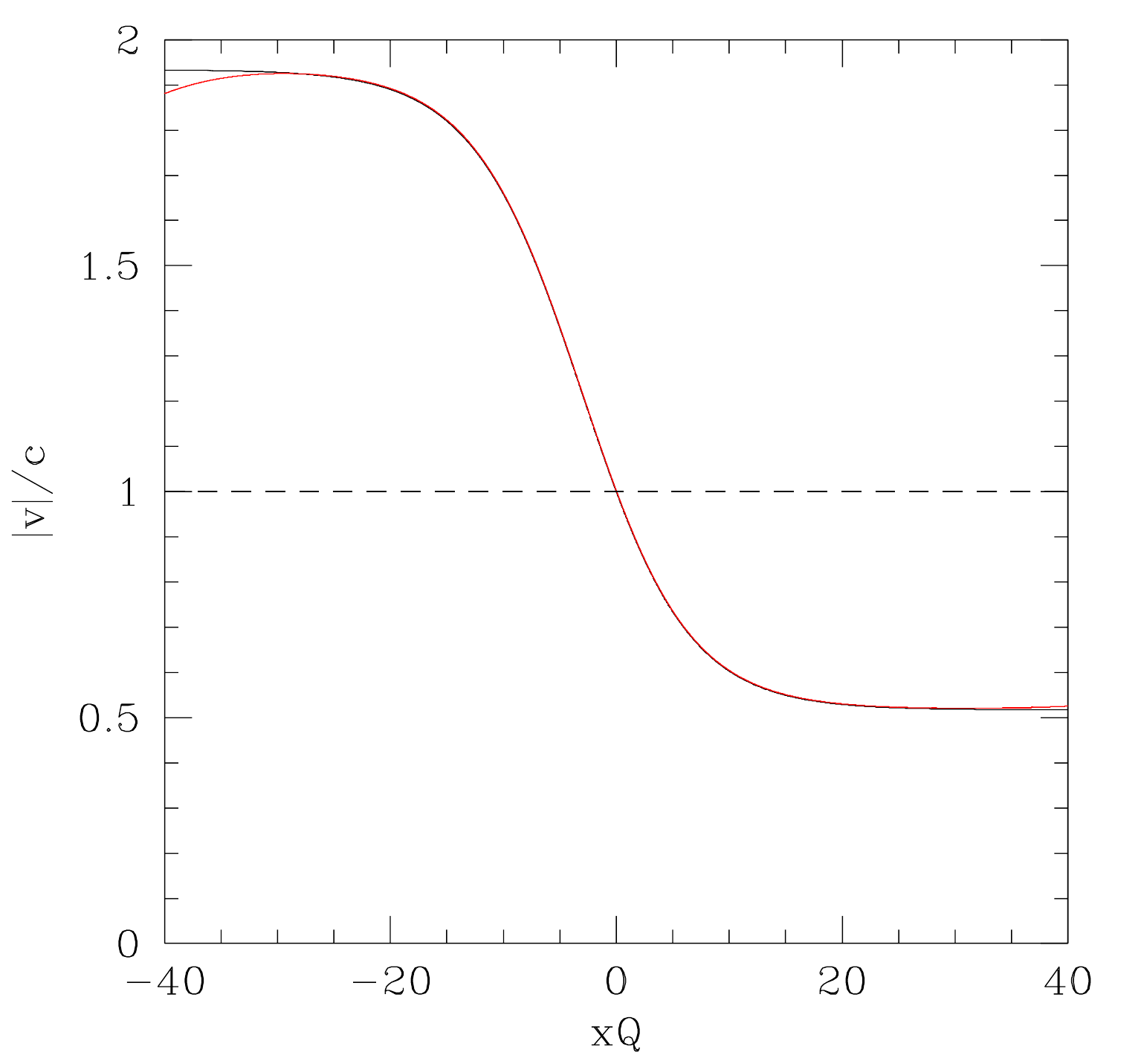}
\caption{Ratio between the absolute value of the fluid velocity and the local sound speed for the same parameter choice as Fig. \ref{fig-evol-a3}. The dashed line shows the ratio before the quench. The analytical stationary solution (black line) is compared to the numerical solution of the Schr\"odinger equation after a time $200\,\frac{m}{\hbar Q^2}$ (red curve).}
\label{fig-evol-a4}
\end{center}
\end{figure}

A comparison between the numerical integration of the Schr\"odinger equation for long times and the asymptotic expressions obtained from Eq. (\ref{hyper}) is shown in Figs. \ref{fig-evol-a1} and \ref{fig-evol-a2} for the density and the velocity profiles, respectively. 
Here the parameter $\alpha$ has been chosen to represent a rather sharp barrier ($\alpha = Q$), while the same comparison for a smoothly varying potential with $\alpha = 0.1\,Q$ is presented in Figs. \ref{fig-evol-a3} and \ref{fig-evol-a4}. 
As for the case of the step, we conclude that the stationary state predicted on the basis of the eigenfunctions (\ref{hyper}) is indeed reached at long times, irrespective of the value of the parameter $\alpha$. An interesting peculiarity of the time evolution in the case of the potential barrier can be readily noticed: in the step case two density modulations originate from the defect (i.e., the waterfall placed at $x=0$) propagating in the upstream and downstream directions; for a barrier, instead, as the supersonic transition sets in, a further soliton-like wave propagating downstream is clearly visible. The soliton velocity is indeed very close to the theoretical expectation $v_{sol} \sim c+|v| = \pi\frac{\hbar Q}{m}$. This observation confirms the analysis, based on the Gross-Pitaevskii dynamics, carried out in Ref. \cite{michel-parentani} where the effect was attributed to the peculiar form of the Bogoliubov excitation spectrum in the supersonic case. Note also that the stationary state properties of the system are visibly different from those of the step potential, while they barely change with $\alpha$ in the range we have examined (modulo a trivial re-scaling of the length unit), showing that, in the case of a potential barrier, the ``smooth limit" is easily achieved also for moderate values of $\alpha/Q$. 

\begin{figure} [ht!]
\begin{center}
\includegraphics[width=7cm]{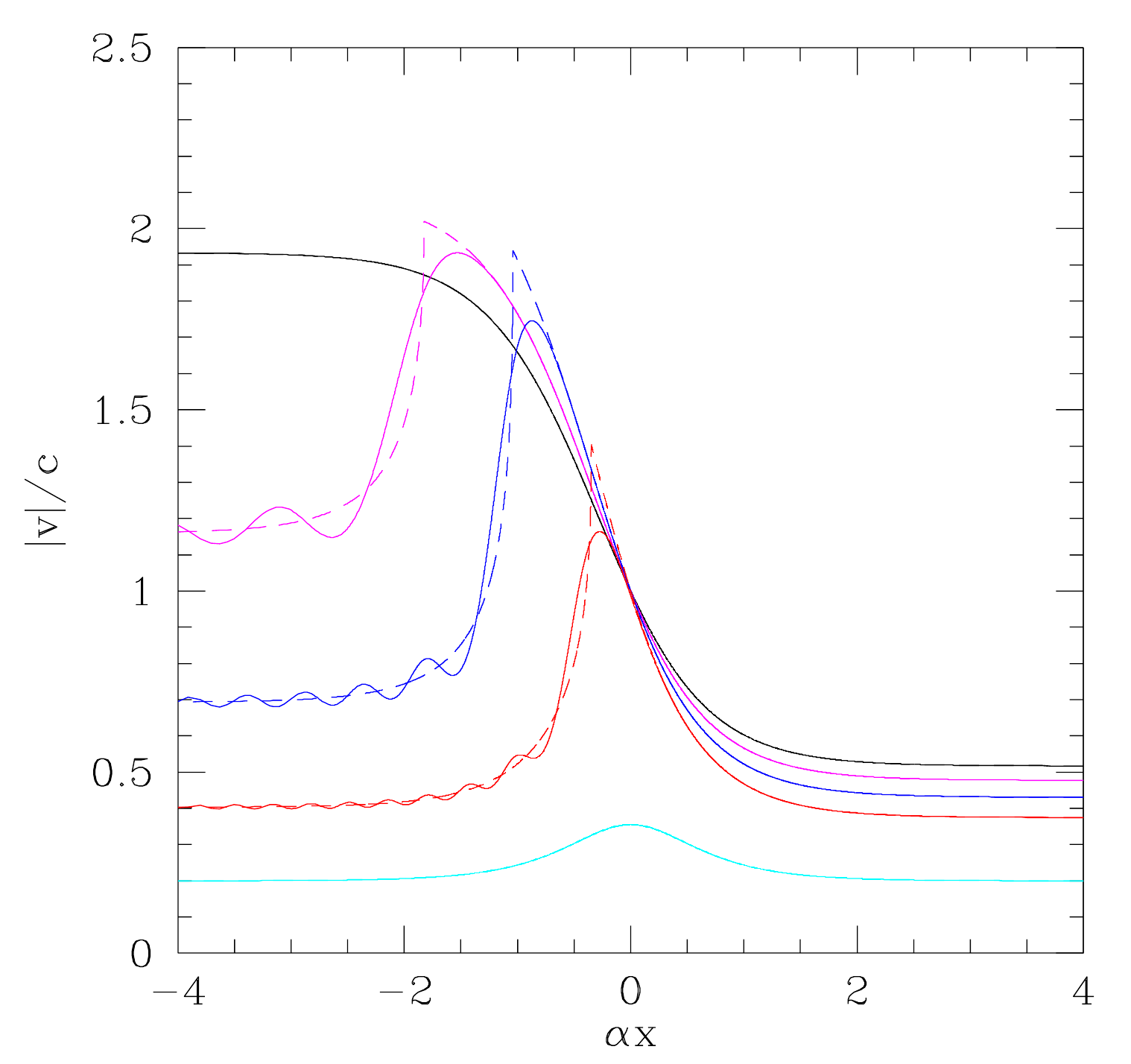}
\caption{Ratio between the absolute value of the fluid velocity and the local sound speed for the stationary state. Here $\alpha=0.1\,Q$, $k_F=\frac{\pi}{2}\,Q$ and $k_0=\frac{5\pi}{10}\,Q$ (i.e. $k_F-k_0=0$, black curve), $k_0=\frac{4\pi}{10}\,Q$ (i.e. $k_F-k_0\sim 0.3 \, Q$, magenta), $k_0=\frac{3\pi}{10}\,Q$ (i.e. $k_F-k_0 \sim 0.6 \, Q$, blue), $k_0=\frac{2\pi}{10}\,Q$ (i.e. $k_F-k_0 \sim 0.9 \, Q$, red), $k_0=\frac{\pi}{10}\,Q$ (i.e. $k_F-k_0 \sim 1.2 \, Q$, cyan). The numerical results are shown in solid lines and they are compared to the analytical expression valid for $\alpha\to 0$ (dashed lines). All cases obey the condition $Q<k_F+k_0$.}
\label{fig-mach}
\end{center}
\end{figure}

Remarkably, in the $\alpha \to 0$ limit, several analytical expressions can be obtained from the asymptotic behavior of the Hypergeometric functions (see Appendix \ref{app:hyper}). The density profile in stationary state is readily evaluated by inserting the results (\ref{asinto1}, \ref{asinto2}) into
\begin{equation}
\rho(x) = \int_{-k_F-k_0}^{k_F-k_0} dk \,\vert \phi_k(x)\vert^2  \, .
\label{rhox}
\end{equation}
It is convenient to define a dimensionless coordinate as
\begin{align}
 \xi&=\tanh (\alpha x), \\ 
\xi_\pm^2&=1-\left( \frac {k_F\pm k_0}Q \right)^2,
\end{align}
The density profile has different analytical expressions in three regimes: 
\begin{itemize}
\item $Q>k_F+k_0$. Here all the fermions have kinetic energy lower than the height of the barrier. Total reflection occurs and the stationary state density profile is formally given by the expression
\begin{align}
 \rho(x)=
\begin{cases}
 \frac Q\pi \sqrt{\xi^2-\xi_-^2} & \mbox{ for } \xi<-\xi_- \\
 \frac Q\pi \sqrt{\xi^2-\xi_+^2} & \mbox{ for } \xi>\xi_+
\end{cases},
\end{align}
while $\rho(x)=0$ elsewhere. The mass flux in this regime vanishes because particles cannot tunnel through the barrier. 
\item The most interesting regime is for $k_F-k_0<Q<k_F+k_0$. Here quantum tunneling occurs and the constant mass flux 
is $j=-\frac{\hbar}{4\pi}\left [ (k_F+k_0)^2-Q^2\right ]$. The density profile is now given by 
\begin{align}
 \rho(x)= \Theta (-\xi_--\xi) \frac Q\pi \sqrt{\xi^2-\xi_-^2}+\frac Q{2\pi} \sqrt{\xi^2-\xi_+^2} +\frac Q{2\pi}\xi,
\end{align}
where the Heaviside function $\Theta(x)=1$ for $x>0$ and vanishes for $x<0$. Note that the density profile is generally
non monotonic. A sonic horizon is always present in this regime.  
\item For $Q<k_F-k_0$ the flow is fully subsonic in the stationary state. The mass flux is  $j=-\frac{\hbar}{\pi}k_F k_0$
and the full density profile preserves the symmetry of the potential: 
\begin{align}
 \rho(x)=\frac Q{2\pi} \left( \sqrt{\xi^2-\xi_-^2}+\sqrt{\xi^2-\xi_+^2} \right).
\end{align}
\end{itemize}
A comparison between the exact solution and the asymptotic analytical results for $\alpha\to 0$ is shown in Fig. \ref{fig-mach} where the local Mach number $\frac{v(x)}{c(x)} =\frac{j}{\pi\hbar\rho(x)^2} $ is displayed for $\frac{\alpha}{Q} =0.1$ and different choices of the parameters $(k_F,k_0)$. 
The analytical expressions are able to capture the essential features of the exact solution, showing that the Mach number develops a maximum in the supersonic region, while downstream the flow may even become subsonic again (blue and red curves), in close analogy with the case of the step potential (see Fig. \ref{fig-vel-tot}). After the peak, the density profile (and therefore also the Mach number) shows oscillations on a scale $x \sim \alpha^{-1}$ which are washed out in the $\alpha\to 0$ limit (dashed lines). The oscillations disappear both in the fully subsonic regime (cyan curve) and for $k_0=k_F$, when the Mach number is always monotonic (black curve). From the analytical expressions we can also evaluate the asymptotic uniform density and velocity in the far upstream region $x\to+\infty$: while in the subsonic case $Q < k_F-k_0$ the asymptotic density and velocity remain unchanged during the quench ($\rho_+=\frac{k_F}{\pi}, v_+=-\frac{\hbar k_0}{m}$), in the interesting regime $k_F-k_0 < Q < k_F+k_0$ they become 
\begin{align}
 \rho_+&= \frac {k_F+k_0+Q}{2\pi},\\
 v_+&= -\frac{\hbar}{2m}\,(k_F+k_0-Q).
\end{align}
Therefore, a hypothetical observer in the far upstream region, unaware of the presence of the potential barrier, would assign an effective value to the key parameters $(k_F,k_0)$ given by 
\begin{align}
 k_F^{eff}&= \frac {k_F+k_0+Q}{2},\\
 k_0^{eff}&= \frac {k_F+k_0-Q}{2}
\end{align}
and then an effective Fermi distribution limited by the two Fermi momenta 
\begin{align}
-k_F^{eff}-k_0^{eff} &= -k_F-k_0, \cr\, 
k_F^{eff}-k_0^{eff} &= Q. 
\label{keff}
\end{align}

At this point, before studying the emergence of the analogue Hawking radiation in the cases presented above, it is important to note how this exact model already displays important differences from the usual semiclassical approach, for both the cases of a waterfall and that of a barrier potential in the $\alpha \to 0$ limit. Indeed, in the presence of a sonic transition, our model shows that the density and the velocities profiles can acquire several different forms in the stationary state; this, though, is profoundly different from the semiclassical approach usually adopted, which, in these cases, only admits monotonic stationary profiles, thus neglecting a whole class of solutions. Here, instead, several different dynamical features which are usually hidden can be investigated.
\section{Analogue Hawking radiation}
At this point, we are ready to investigate the occurrence of thermal phonons emerging from the sonic horizon in the stationary state of the model. 
We stress that, while in the gravitational framework the background curved metric originates from the presence of a black hole which is unrelated to the quantum field giving rise to the Hawking radiation, in the condensed matter analogy, both the effective metric and the scalar field are expressions of the same quantum fluid: the flowing Bose gas. The phonon field, which plays the role of the quantum scalar field in this framework, can indeed be considered decoupled from the fluid flow only if we limit our analysis to low energy, where the elementary excitations are known to behave as free quantum quasi-particles \cite{pines}. 

The energy density of a one-dimensional uniform Bose fluid in the presence of a thermal phonon branch is known to exhibit an additive contribution of the form 
\begin{equation}
E_{ph}(T) = \int_{0}^\infty \frac{dk}{2\pi}\,\frac{\hbar c k }{e^{\beta\hbar c k} -1} =
\frac{\pi}{12} \, \frac{(k_BT)^2}{\hbar c}
\label{ebose}
\end{equation}
where the usual phonon dispersion relation $\omega=c\,k$ has been used. Switching to the representation of the HCB in terms of spinless fermions, the same fluid can be described in terms of an effective Fermi-Dirac distribution
\begin{equation}
f_{\text{FD}}(k) = \frac{1}{1+e^{\beta(\epsilon_k - \mu)}} \, ,
\label{dirac}
\end{equation}
leading to a thermal contribution to the energy density
\begin{eqnarray}
\label{efermi}
E(T) &=& \int_{-\infty}^\infty \frac{dk}{2\pi}\,f_{\text{FD}}(k)\,\epsilon_k  \\
&=& E(0) + \frac{\pi}{6} \, \frac{m}{\hbar^2 k_F}\,(k_BT)^2 +O(T^4) \nonumber \, .
\end{eqnarray}
The first term is the ground state energy of the Fermi gas, while the second equals the thermal energy density of the two phonon branches (one for each Fermi point), as can be easily checked recalling that the sound velocity in a one-dimensional Fermi gas (or a HCB fluid) coincides with the Fermi velocity, $c=v_F = \frac{\hbar k_F}{m}$. This simple observation shows that, in one dimension, there is a one-to-one correspondence between the quantitative description of the one-dimensional HCB system in terms of a phonon gas and that of a fluid of fermionic particles, if only low energy excitations are present. 

Moreover, if phonons originate from the sonic horizon and we study the flow in the far upstream region $x\to +\infty$, only quasiparticles with positive wavevectors $k>0$ will appear while left moving particles (i.e. with $k<0$) would be unaffected by the Hawking mechanism and remain at zero temperature. Therefore, we are led to a very specific expectation for the effective momentum distribution of the fermions in the region $x\to +\infty$ if analogue Hawking emission occurs: all local physical observables, in fact, should appear as if the Fermi gas was characterized by a momentum distribution of the form (\ref{dirac}) with $\beta = (k_BT_H)^{-1}$ for $k > 0 $ and $\beta = \infty$ for $ k < 0 $. 

Finally, another signature of the particle emission at the horizon due to the Hawking process is related to the existence of quantum correlations between the upstream and the downstream regions \cite{iacopo}. Physically, these correlations originate from the creation of entangled phonon pairs at the horizon propagating in opposite directions. In our microscopic model, we can check also this prediction by evaluating the density-density correlations in the stationary state:
\begin{eqnarray}
\label{corre}
h(x,x^\prime) &=& \frac{\langle \hat\rho(x,t)\hat\rho(x^\prime,t)\rangle}{\rho(x,t)\rho(x^\prime,t)} -1
\\
&=& -\,\frac{\left \vert \int_{-k_F-k_0}^{k_F-k_0} dk\,
\psi_k^*(x,t)\,\psi_k(x^\prime,t) \right\vert^2}{\rho(x)\rho(x^\prime)} \nonumber  \, ,
\end{eqnarray} 
where the local density $\rho(x)=\langle\hat\rho(x,t)\rangle = \langle \hat\psi^\dagger(x,t)\hat\psi(x,t)\rangle$ is defined in Eq. (\ref{dens}). We now turn to the evaluation of these properties for the representative choices of potentials previously introduced.

\subsection{Step potential}

As already mentioned, in the far upstream region $x \to+\infty$ the asymptotic wavefunction $\psi_k(x,t)$ (\ref{asy}) simplifies and it becomes the superposition of an incident and a reflected plane wave (this is true regardless of the form of the potential). Omitting the overall time dependent phase factor, the asymptotic expression of $\psi_k(x,t)$ for a step potential is: 
\begin{equation}
\begin{cases} 
\frac{1}{\sqrt{2\pi}}\,\frac{2k}{k+\sqrt{k^2-Q^2}}\, e^{ix\sqrt{k^2-Q^2}} 
& k>Q  \\
\frac{1}{\sqrt{2\pi}}\,\left [ e^{ikx} - \left(\sqrt{1+\frac{k^2}{Q^2}} + \frac{k}{Q} \right)^2 \,e^{-ikx}
\right]
& k<0 \nonumber
\end{cases}
\end{equation}
while for $0<k \le Q$ the wavefunction is exponentially small at $x\to +\infty$. When the quantum average of a physical quantity is evaluated starting from this expression, the asymptotic result at large $x$ can be formally written according to Eqs. (\ref{1b},\ref{2b}). Disregarding the interference term between the two counter-propagating waves whose contribution vanishes for $x\to +\infty$, the limiting form coincides with that of a uniform free Fermi gas characterized by an effective momentum distribution $f(k)$ given by: 
\begin{equation}
\begin{cases}
\,\quad 1 &   \mbox{for } -k_F-k_0<k< K  \\
\left(\frac{p-k}{Q} \right)^4 
&  \mbox{for } K < k<k_F+k_0
\end{cases}
\nonumber
\end{equation}
where we have defined $p=\sqrt{k^2+Q^2}$, $K=\sqrt{(k_F-k_0)^2-Q^2}$ for $k_F-k_0 > Q$ and $K=0$ elsewhere. This momentum distribution displays a tail at positive wavevectors, denoting the presence of quasiparticles (phonons) traveling upstream in the stationary state. The analytic form of the tail coincides with the reflection coefficient of the external potential $V(x)$.  According to this expression, for $k_F-k_0 > Q$ the flow is purely subsonic, $K>0$ and the momentum distribution preserves the sharp discontinuity at $k=K$. Conversely, when a supersonic transition is present, $K=0$ and the two branches of $f(k)$ join smoothly at $k=0$.  Although the qualitative behavior of the system conforms to the expectations based on the gravitational analogy, the quantitative details do not: a phonon flux is still present even in the absence of a sonic horizon and, most importantly, the ``Hawking-like" radiation is never thermal, because the effective distribution differs from  the Fermi-Dirac form (\ref{dirac}). This result denotes a failure of the gravitational analogy, which, being based on semiclassical arguments,
is not expected to faithfully represent the actual behavior of the model when a rapidly varying external potential, like a step, is switched on. 

This analysis can be generalized for the case of a smooth step of the form $V(x)=\frac{V_0}{2}\left [ 1+ \tanh\alpha x\right ]$ where the parameter $\alpha$ controls the sharpness of the potential and the discontinuous step is recovered for $\alpha\to\infty$. By defining $V_0=\frac{\hbar^2Q^2}{2m}$ and calculating the reflection coefficient for this case \cite{landau}, the effective momentum distribution turns out to be 
\begin{equation}
\begin{cases}
\quad\qquad  1 &   \mbox{for } -k_F-k_0<k< K  \\
\left[
\frac{\sinh\left ( \frac{\pi}{2\alpha}(p-k)\right)}
{\sinh\left ( \frac{\pi}{2\alpha}(p+k)\right)}
 \right]^2 
&  \mbox{for } K < k<k_F+k_0
\end{cases}
\nonumber
\end{equation}
where $p$ and $K$ have been previously defined. We see that the general features of this effective momentum distribution remain unchanged for a smooth step, even in the limit $\alpha\to 0$, when the phonon tail reduces to $e^{-\frac{2\pi k}{\alpha}}$. Thus, in the case of an external potential which has a step form (smooth or sharp) we have emission of phonons in the $x \to +\infty$ region from the horizon but the spectrum is never thermal. Furthermore, the emission persists even if a supersonic transition is absent and disappears only for $\alpha\to 0$.

\begin{figure}[ht!]
\begin{center}
\includegraphics[width=7cm]{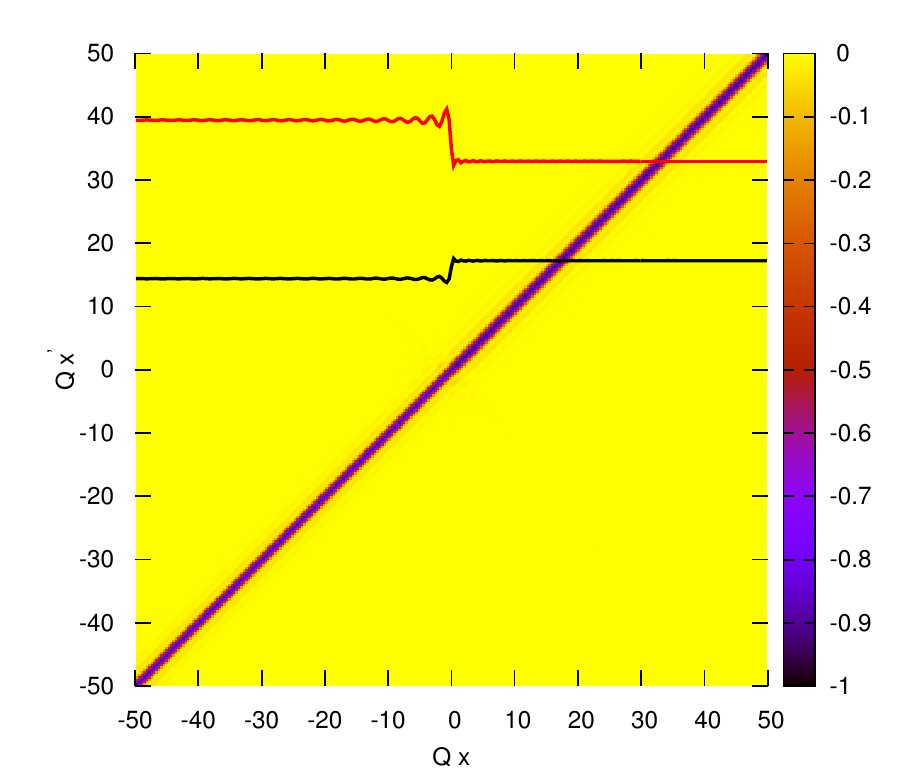}
\caption{Color density plot of the correlation function (\ref{corre}) for a step potential with $k_F=\frac{3\pi}{5}\,Q$ and $k_0=\frac{\pi}{5}\,Q$. The absolute value of the fluid and sound velocities (black and red curves, respectively) are reported in arbitrary units to illustrate the absence of a sonic horizon.}
\label{fig-corr-sub}
\end{center}
\end{figure}
\begin{figure}[ht!]
\begin{center}
\includegraphics[width=7cm]{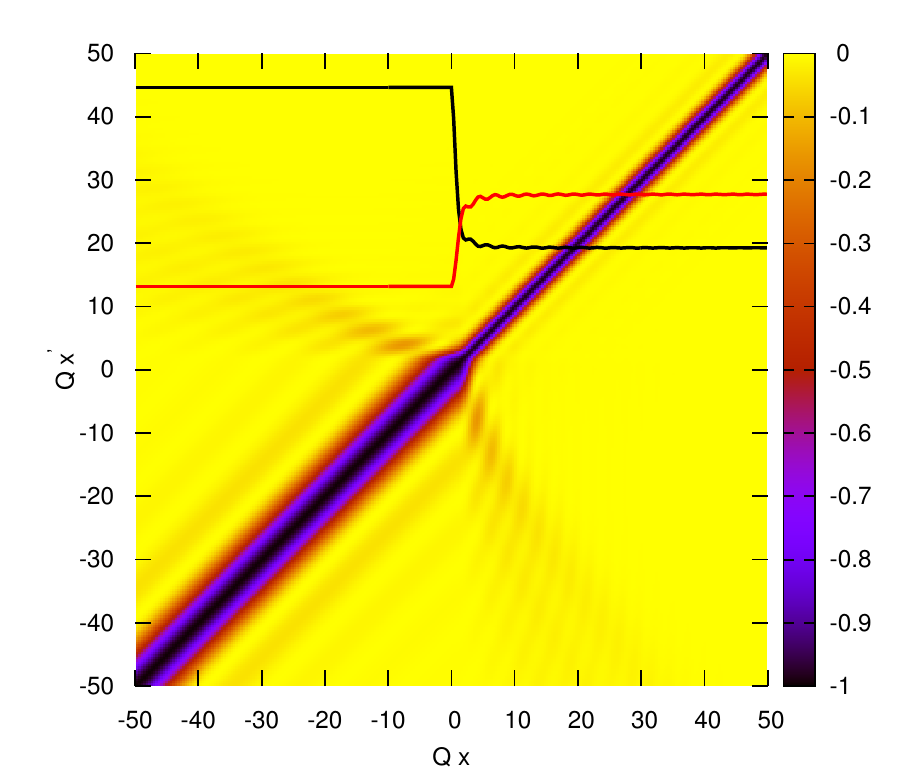}
\caption{Color density plot of the correlation function (\ref{corre}) for a step potential with $k_F=k_0=\frac{\pi}{5}\,Q$. The absolute value of the fluid and sound velocities (black and red curves, respectively) are also reported in arbitrary units to indicate the position of the sonic horizon.}
\label{fig-corr-super}
\end{center}
\end{figure}

The stationary state density correlations (\ref{corre}) can be easily computed starting from the asymptotic form of the wavefunctions (\ref{asy}). It is interesting to compare the results in the fully subsonic regime ($k_F-k_0 > Q$) and the ones in the supersonic regime ($0< k_F-k_0 < Q$). In Figs. \ref{fig-corr-sub} and \ref{fig-corr-super} the results are shown in the case of a sharp step for two representative choices of the parameters. 
The presence of weak, but clearly-visible, correlations across the sonic horizon in Fig. \ref{fig-corr-super} confirms the commonly accepted picture: phonon pairs created at the horizon and propagating upstream and downstream give rise to density correlations between the inner and the outer regions. What is more interesting, though, is that this pattern appears only for a supersonic transition (in contrast with the particle emission which is always present) but it emerges even if the radiation emitted is not thermal.

\subsection{Repulsive barrier}
For the repulsive barrier (\ref{cosh}) the asymptotic wave function $\psi_k$ in the upstream region $x\to+\infty$ takes the form
\begin{equation}
\psi_k(x) =
\begin{cases}
\frac{1}{\sqrt {2\pi}}\,T_k \,e^{ikx} & \mbox{for } k>0\\
\frac{1}{\sqrt {2\pi}}\,\left [ 
e^{ikx}+R_k \,e^{-ikx}\right ] & \mbox{for } k<0
\end{cases},
\end{equation}
where, again, we omitted an overall time dependent phase factor, and
\begin{align}
\label{trans}
T_k&=\frac {\Gamma(\frac 12-i\frac {|k|+Q}{\alpha}) \Gamma(\frac 12
- i\frac {|k|-Q}{\alpha})}{\Gamma(1-i\frac {|k|}{\alpha})\Gamma(-i\frac {|k|}{\alpha})},\\
R_k&=\frac {\Gamma(i\frac {|k|}{\alpha})
\Gamma(\frac 12 -i\frac {|k|+Q}{\alpha}) 
\Gamma(\frac 12-i\frac {|k|-Q}{\alpha})}
{\Gamma(-i\frac {|k|}{\alpha})\Gamma(\frac 12-i\frac {Q}{\alpha}) \Gamma(\frac 12+i\frac {Q}{\alpha})},
\label{ref}
\end{align}
are the transmission and reflection amplitudes. As before, we can regard this asymptotic form as an equivalent Fermi gas with effective momentum distribution $f(k)$ given by
\begin{align}
\begin{cases}
 1 & \mbox{for } -k_F-k_0 < k <k_F-k_0 \\
 |R_k|^2  & \mbox{for } k_F-k_0 < k <k_F+k_0
\end{cases} \, ,
\label{effect}
\end{align}
with
\begin{align}
|R_k|^2= \frac {1+\cosh (2\pi Q/\alpha)}{\cosh (2\pi Q/\alpha)+\cosh (2\pi k/\alpha)}.
\end{align}
Note that $f(k)$ is continuous in $k=k_F-k_0$ only for $k_F=k_0$. For a generic value of $\frac{\alpha}{Q}$, the effective momentum distribution (\ref{effect}) differs from the expected Fermi-Dirac form (\ref{dirac}), reflecting the coupling between the quasiparticles (phonons) and the underlying metric (flowing Bose gas). Only for $\frac{\alpha}{Q}\to 0$, i.e. for very smooth barriers, phonons are excited at extremely low energies and quasiparticles behave as a free scalar field. In this limit, the effective momentum distribution acquires the suggestive form 
\begin{align}
f(k) \simeq 
\frac 1{1+e^{\frac {2\pi}\alpha (k-Q)}} 
\label{effective}
\end{align}
for $k_F-k_0 < k <k_F+k_0$, while $f(k)=1$ for $-k_F-k_0 < k <k_F-k_0$. This distribution describes a Fermi gas with a sharp jump at the left Fermi point $-k_F-k_0$, while a tail appears for $k>k_F-k_0>0$, indicating the presence of excited phonons in the HCB fluid. First, we note that both in the fully subsonic ($k_F-k_0 >Q$) and in the fully supersonic regime ($Q> k_F+k_0$) the effective Fermi point $k=Q$ lies outside the interval where Eq. (\ref{effective}) holds. Then, for $\alpha\to 0$, $f(k)$ either vanishes (for $k_F-k_0 >Q$) or is identically equal to one (for $Q> k_F+k_0$). In both cases, for $\alpha \to 0$ the effective distribution coincides with the standard zero temperature result and no phonon flux is present in the HCB fluid at $x\to\infty$. Instead, in the interesting regime where the sonic horizon forms ($k_F-k_0<Q<k_F+k_0$), we can quantitatively match the result (\ref{effective}) with the expected Fermi-Dirac distribution (\ref{dirac}) at finite temperature. Indeed, let us start from Eq. (\ref{keff}) and linearize the energy spectrum near the effective Fermi point $k=Q$, corresponding to the chemical potential $\mu=\epsilon_Q$, giving the quasiparticle dispersion at low energy: 
\begin{align}
 \epsilon_k-\mu \simeq \hbar v_{qp} \,(k-Q),
\end{align}
with quasiparticle velocity $v_{qp}=\hbar Q/m$. Inserting such a form into the Fermi distribution (\ref{dirac}), we obtain the expression Eq. (\ref{effective}) if the effective temperature $T_H$ is given by 
\begin{align}
 k_B T_H=\alpha \frac {\hbar^2 Q}{2\pi m}.
\label{th}
\end{align}
Therefore, as previously discussed, an observer at $x\to+\infty$ will detect a phonon field at the temperature $T_H$ on top of the flowing HCB fluid. As we will see, expression (\ref{th}) coincides with the Hawking temperature predicted by the analogue gravity description. 
\begin{figure}
\begin{center}
\includegraphics[width=7cm]{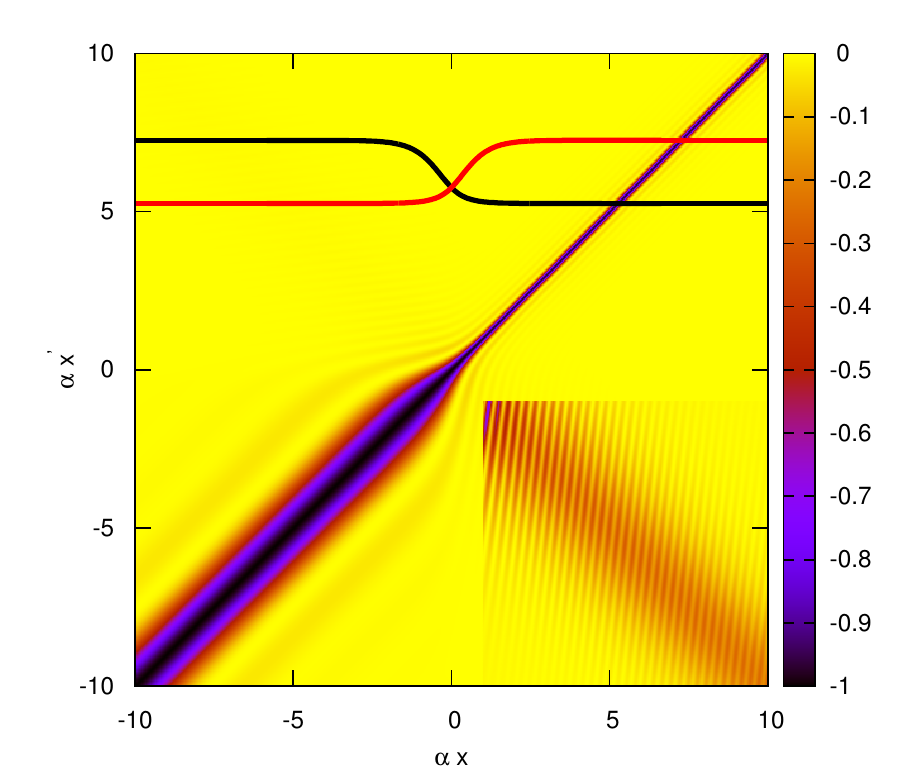}
\caption{Colour density plot of the correlation function (\ref{corre}) for a repulsive barrier with
$\alpha =0.1\,Q$ and $k_F=k_0=\frac{\pi}{5}\,Q$. The absolute value of the fluid and sound velocities (black and red curves, respectively) are also reported in arbitrary units to indicate the position of the sonic horizon. In the bottom-right corner of the figure a blow-up (by a factor $50$) of the off-diagonal correlations is shown.}
\label{fig-corr-cosh}
\end{center}
\end{figure}
We can also evaluate the density correlations (\ref{corre}) on the basis of the analytical form of the scattering states (\ref{hyper}). A representative result is shown in Fig. \ref{fig-corr-cosh} for $\alpha=0.1\,Q$ and $k_F=k_0=\frac{\pi}{5}\,Q$. The fluid velocity for the stationary state is also reported (black line) together with the local sound velocity (red line). In the bottom-right corner a blow-up of the off-diagonal correlations is displayed in order to appreciate the presence of the weak signal related to the emergence of the Hawking radiation.

An analytical understanding of the density correlations is possible in the asymptotic region $(x,x^\prime)\to\pm\infty$ in the limit $\alpha\to 0$. Focusing on the interesting range of parameters $k_F-k_0 < Q < k_F+k_0$ where the horizon is present and taking advantage of the explicit expressions of the transmission and reflection coefficients (\ref{trans}, \ref{ref}), the asymptotic correlations can be written in terms of $k^\pm = k_F\pm k_0$ as 
\begin{itemize}
\item $x\to +\infty ,\,x^\prime\to +\infty$. Here the fluid is subsonic and the correlations become translationally invariant asymptotically. Setting $s=x-x^\prime$ we obtain, to leading order in $\alpha$, 
\begin{equation}
h_+(s) = - \left [ \frac{\sin\frac{k^++Q}{2}s}{\frac{(k^++Q)\, s}{2}}\right ]^2 \, .
\end{equation}
\item $x\to -\infty , \,x^\prime\to -\infty$. In the downstream region density correlations show a more complex structure as a function 
of $s=x-x^\prime$
\begin{equation}
h_-(s) = -\left \vert \frac{\sin k^-s + e^{i\frac{k^++Q}{2}s}\,\sin\frac{k^+-Q}{2}s
}{\frac{(k^++2k^--Q)\, s}{2}}\right \vert^2 .
\end{equation}
\item $x\to \pm\infty , \,x^\prime\to \mp\infty$. To leading order in $\alpha$, the off-diagonal correlations identically vanish far from the horizon. However, a non trivial result appears to second order, proving that density correlations between specular points in the sub- and super-sonic regions persist even at large distances from the barrier. Setting $u=x+x^\prime$, the  asymptotic form of the correlation function is 
\begin{equation}
h_\pm(u) = -\frac{ \alpha^2\, \left [ \cosh\frac{\alpha u}{2}\right]^{-2}}
{4\,(k^++Q)\,(k^++2k^--Q)} .
\end{equation}
Remarkably, this exact result for the Tonks-Girardeau gas reproduces the semiclassical prediction of Ref. \cite{balbinot} 
obtained in a mean field model where the effective coupling $g$ of the Gross-Pitaevskii equation acquires different values in the upstream and downstream regions.  
\end{itemize}


\section{Semiclassical analysis}
\label{semiclassical}
A semiclassical description of the Tonks-Girardeau gas can be obtained considering a Bose fluid in the second quantization formalism, described by the Hamiltonian
\begin{align}
 \hat H=\int dx \left\{ \hat \psi^\dagger (x) \left( -\frac {\hbar^2}{2m} \frac {d^2}{dx^2}\right) \hat \psi (x) +W \right\},
\end{align}
where $W=W_c+W_e$ is the sum of a contact term
\begin{align}
 W_c=\frac g\nu \left( \hat \psi^\dagger (x) \right)^\nu \left( \hat \psi(x) \right)^\nu 
\label{contact}
\end{align}
and the interaction with an external potential
\begin{align}
 W_e=(V(x)-\mu) \hat \psi^\dagger (x) \hat \psi(x).
\end{align}
For future reference, we consider a generic $\nu$-body interaction so that the usual pair potential corresponds to 
the choice $\nu=2$ in Eq. (\ref{contact}). The bosonic field operator $\hat \psi(x)$ satisfies the canonical commutation relations
\begin{align}\label{ccr}
 [\hat \psi(x), \hat \psi^\dagger(x')]=\delta(x-x').
\end{align}
We can decompose the field operator as a sum of a background configuration, described by a complex function $\psi(x)$, and a quantum perturbation so that
\begin{align}
 \hat \psi(x)=\psi(x)+\delta\hat \psi(x),
\end{align}
provided that $\delta\hat\psi(x)$ satisfies the relations (\ref{ccr}). Inserting in the Hamiltonian and keeping terms up to order two in the perturbation, we get $\hat H=E+\hat H_1+\hat H_2$, with
\begin{align}
& \hat H_1=\int dx \, \delta \hat \psi^\dagger  h_1\, \psi+h.c.\\
& \hat H_2=\int dx \left\{ \delta\hat \psi^\dagger h_2\delta \hat\psi\right. +\cr
& +\left.\frac g2 (\nu-1) |\psi|^{2\nu-4} \left( \psi^2 \delta\hat \psi^2+h.c. \right)\right\},
\label{H2}
\end{align}
where $E$ is the reference ``classical" energy and 
\begin{align}
\label{h1}
 h_1(x)=-\frac {\hbar^2}{2m} \frac {d^2}{dx^2}+g|\psi(x)|^{2\nu-2}+V(x)-\mu,\\
 h_2(x)=-\frac {\hbar^2}{2m} \frac {d^2}{dx^2}+g\nu|\psi(x)|^{2\nu-2}+V(x)-\mu
\label{h2}
\end{align}
are the first quantization effective Hamiltonians. The stationary background configuration $\psi(x)$ is chosen so that the first order contribution $\hat H_1$ identically vanishes:
\begin{align}\label{background}
 h_1(x)\psi(x)=0.
\end{align}
This non-linear differential equation defines the stationary solution in the semiclassical approximation and corresponds, for $\nu=2$, to the known Gross-Pitaevskii equation for the condensate wavefunction. While the obvious choice $\nu=2$ corresponds to the physical many body Hamiltonian (\ref{h}), it has been shown \cite{menotti-stringari,seir,alb} that the semiclassical approximation reproduces the exact spectrum of the strongly interacting Tonks-Girardeau gas in one dimension for the alternate choice: 
\begin{align}
 \nu=3 \qquad\ \mbox{ and } \qquad\ g=\frac {\hbar^2 \pi^2}{2m}.
\end{align}
These are the values we will consider from now on.

Note that this change in the dynamical equations does not affect the validity of the gravitational analogy. It is possible, in fact, to demonstrate the existence of an effective acoustic metric also starting from this generalized Gross-Pitaevskii equation for any $\nu$. The procedure implies the usual derivation of the acoustic metric \cite{garay} which turns out to have the canonical form, apart from the conformal factor.
\subsection{The stationary configuration}
We will now assume an external smooth potential of the form
\begin{align}
 V(x)=U(\alpha x)
\end{align}
in the limit $\alpha\to0$. It is convenient to introduce the dimensionless variable $z=\alpha x$. In order to solve the generalized Gross-Pitaevskii equation (\ref{background}) with $\nu$ and $g$ just introduced, we make the Ansatz 
\begin{align}\label{ansatz}
 \psi(x)=A(z) e^{i\phi(z)/\alpha}.
\end{align}
Inserting this form in Eq. (\ref{background}), at first order in $\alpha$ we get the equation
\begin{align}
2A'(z) \phi'(z)+A(z) \phi''(z)=0,
\end{align}
where a prime indicate derivative w.r.t. $z$. This expresses mass conservation, which, for a stationary flow in one dimension, leads to the constancy of the mass current
\begin{align}\label{current}
 j=\hbar A^2(z)\phi'(z).
\end{align}
Solving for $\phi'(z)$ and substituting in the zeroth order equation obtained from (\ref{background}) and (\ref{ansatz}), we get
\begin{align}\label{amplitude}
 A^4(z)=\frac {\mu-U(z)\pm \sqrt {(\mu-U(z))^2-\sigma^2}}{2g},
\end{align}
with 
\begin{align}
\sigma=\frac {\pi \hbar j}m.
\end{align}
The chemical potential must therefore satisfy the constraint
\begin{align}
 \mu> U(z)+|\sigma| \qquad\ \forall z,
\end{align}
which allows for two solutions for each compatible choice of $\mu$ and $j$. Recalling that the local velocity $v(z)$ and the sound velocity $c(z)$ for a Tonks gas are
\begin{align}
 v(z)=\frac {j}{m\rho(z)}, \qquad c(z)=\frac {\pi \hbar \rho(z)}m,
\end{align}
with equilibrium density $\rho(z)=A^2(z)$, we get for the Mach number $\beta=v/c$:
\begin{align}\label{Mach}
 \beta(z)=\frac \sigma{\mu-U(z)\pm \sqrt {(\mu-U(z))^2-\sigma^2}}.
\end{align}
The upper and lower signs correspond to a subsonic and a supersonic velocity profile respectively, and no transition (horizon) appears in general, for any potential. The only wavefunction describing a sonic transition is obtained by matching the two solutions at a point $z_0$ where $\mu=U(z_0)+|\sigma|$. In order to join the solutions keeping $\beta(z)$ real, $z_0$ must be a maximum for the potential: $U(z_0)=U_{max}$. 

We can now compare the resulting stationary state of the semiclassical solution with the exact one in the two cases described in the previous Sections.
\subsubsection{Step potential}
Let us consider first a smooth step potential of the form
\begin{align}
U(z)=\frac{U_0}{2}\left [ 1+ \tanh z\right ], \quad\ U_0=\frac {\hbar^2 Q^2}{2m}
\end{align}
in the limit $\alpha\to 0$.
In this case there are no local maxima and, therefore, it is impossible to match the two branches; thus, we are left with either a fully supersonic 
or a fully subsonic regime. 

When $Q>k_F-k_0$, in the exact solution the Mach number approaches $1$ when $x\to+\infty$. Therefore, from (\ref{Mach}), we have to set $\mu=U_0+\sigma$. Parametrizing the mass flux in terms of $k_F$ and $k_0$ as $j=-\frac {\hbar}{4\pi} (k_F+k_0)^2$, we obtain the asymptotic values of the density profile at $z\to \pm\infty$: 
\begin{align}
 & \rho_+=\frac {k_F+k_0}{2\pi}, \\
 & \rho_-=\frac {\sqrt {(k_F+k_0)^2+Q^2}\pm Q}{2\pi},
\end{align}
where $\pm$ in the second formula correspond to the upper or the lower sign in (\ref{amplitude}). The semiclassical solution reproduces correctly the exact asymptotic density at $+\infty$, while at $-\infty$ it gives the correct profile only for $k_F- k_0=0$ (choosing the lower sign) or $k_F-k_0=Q$ (choosing the upper sign).\\
If $Q<k_F-k_0$ the exact solution is subsonic and we have to select the upper sign in (\ref{amplitude}) while $\mu$ can always be chosen so that the semiclassical solution fully reproduces the exact density profile.

It is important to note that the absence of a sonic horizon for a waterfall potential in the $\alpha\to 0$ limit is not a peculiarity caused by the slight modifications introduced in the Gross-Pitaevskii equation. Indeed, it can be demonstrated that also the canonical Gross-Pitaevskii equation (i.e., setting  $\nu=2$ and $g$ the usual interaction parameter) does not admit a sonic transition for any choice of the parameters.

\subsubsection{Repulsive barrier}
Consider a barrier displaying a unique maximum in $z_0$: 
\begin{align}
 U_{max}=U(z_0)=\frac {\hbar^2 Q^2}{2m}
\end{align}
and choose $\mu=\frac {\hbar^2 Q^2}{2m}+|\sigma|$. We can then take the lower sign for $z<z_0$ and the upper sign for $z>z_0$ to get a solution that passes from a supersonic regime on the left of $z_0$ to a subsonic regime on the right. Let us compare this solution with the exact expressions for the potential (\ref{cosh}) previously discussed. The exact solution shows that a subsonic/supersonic transition is present for $k_F-k_0<Q<k_F+k_0$. Choosing $Q$ in this range, we parametrize the mass current in terms of $k_F$ and $k_0$ as $j=\frac {\hbar}{4\pi} [(k_F+k_0)^2-Q^2]$. The density profile in semiclassical approximation $\rho(x)=A(z)^2$ is monotonic and its asymptotic values at $z\to \pm \infty$ are 
\begin{align}
 \rho_\pm =\sqrt {\frac {|j|}{\pi \hbar}} \left( \frac {\mu}{|\sigma| \pm \sqrt {\left( \frac \mu{\sigma}\right)^2-1}} \right)^{\frac 12},
\end{align}
which for the given values of $\mu$ and $j$ give
\begin{align}
 \rho_\pm=\frac {k_0+k_F\pm Q}{2\pi}.
\end{align}
For a  barrier of the form (\ref{cosh}), $\rho_+$ coincides with the exact value, while $\rho_-$ is correct only for $k_F=k_0$, when the exact density profile is monotonic. Note that, in this particular case, not only the asymptotic values but also the full density profile coincides with the exact one displaying a sonic horizon at $z=0$. For future reference we note that the slope of the relative velocity at the horizon $\kappa = \frac{d}{dz} (c(z)+v(z))\vert_{z=0}$ is $\kappa=\frac{\hbar Q}{m}$. It is interesting to consider also the fully subsonic case $Q<k_F-k_0$, where no matching procedure is required. Correspondingly, we choose the upper sign in (\ref{amplitude}) and set $|j|=\frac {\hbar k_F k_0}\pi$. Now $\mu$ is unconstrained and if we parametrize it as $\mu = \frac{\hbar^2}{2m} (k_F^2+k_0^2)$, 
we get that the exact density profile is reproduced by the semiclassical solution for any $k_0<k_F$.

Remarkably, in the fully subsonic case, for both the barrier and the step potential, the exact density profile in the $\alpha\to 0$ limit is always reproduced by the semiclassical one. Instead, if a horizon is present, the semiclassical approximation is correct only for strictly monotonic velocity profiles, i.e. for a very specific (fine tuned) choice of the initial velocity corresponding to a sonic flow. \\
Finally, it is worth to mention that we have just looked for the stationary solutions of the semiclassical equations but we have not checked whether the semiclassical dynamics does indeed drive the system towards such a solution. In the Heisenberg formalism the field operator $\hat \psi(x,t)$ evolves in time according to
\begin{align}
 i\hbar \frac {\partial \hat \psi(x,t)}{\partial t}=-[\hat H,\hat \psi(x,t)],
\end{align}
which, after separating the background configuration from the quantum fluctuations, gives
\begin{align}
\label{eqpsi}
&  i\hbar \frac {\partial \psi(x,t)}{\partial t}=h_1(x,t) \psi(x,t), \\
&  i\hbar \frac {\partial \delta \hat \psi(x,t)}{\partial t}=-[\hat H_2, \delta\hat \psi(x,t)],
\label{eqdeltapsi}
\end{align}
where $h_1(x,t)$ and $\hat H_2(t)$ are formally given by Eqs. (\ref{H2}-\ref{h2}) evaluated in terms of the evolving condensate wavefunction $\psi(x,t)$. Assuming a uniform condensate $\psi(x)=\sqrt {\frac {k_F}\pi}e^{-ik_0x}$ as initial condition, in the $\alpha\to 0$ limit we can look for a solution of the form
\begin{align}
 \psi(x,t)=\sqrt {\rho(z,t)} \, e^{\frac i\alpha \phi(z,\tau)},
\end{align}
expressed in terms of the dimensionless variables
\begin{align}
 & z=\alpha x, \qquad\ \tau =\frac {\hbar Q}m \alpha t. \label{dimensionless}
\end{align}
The equations of motion for the background field in presence of the barrier (\ref{cosh}) now become
\begin{align}
 Q \frac {\partial \rho}{\partial \tau} &=-\frac {\partial}{\partial z} \left[ \rho \phi' \right],\label{density}\\
 Q \frac {\partial \phi'}{\partial \tau} &= -\frac 12 \frac {\partial}{\partial z} \left[ \phi^{\prime2} +\pi^2 \rho^2+\frac {Q^2}{\cosh^2 z} \right].  \label{phase}
\end{align}
\begin{figure}[ht!]
\begin{center}
\includegraphics[width=7cm]{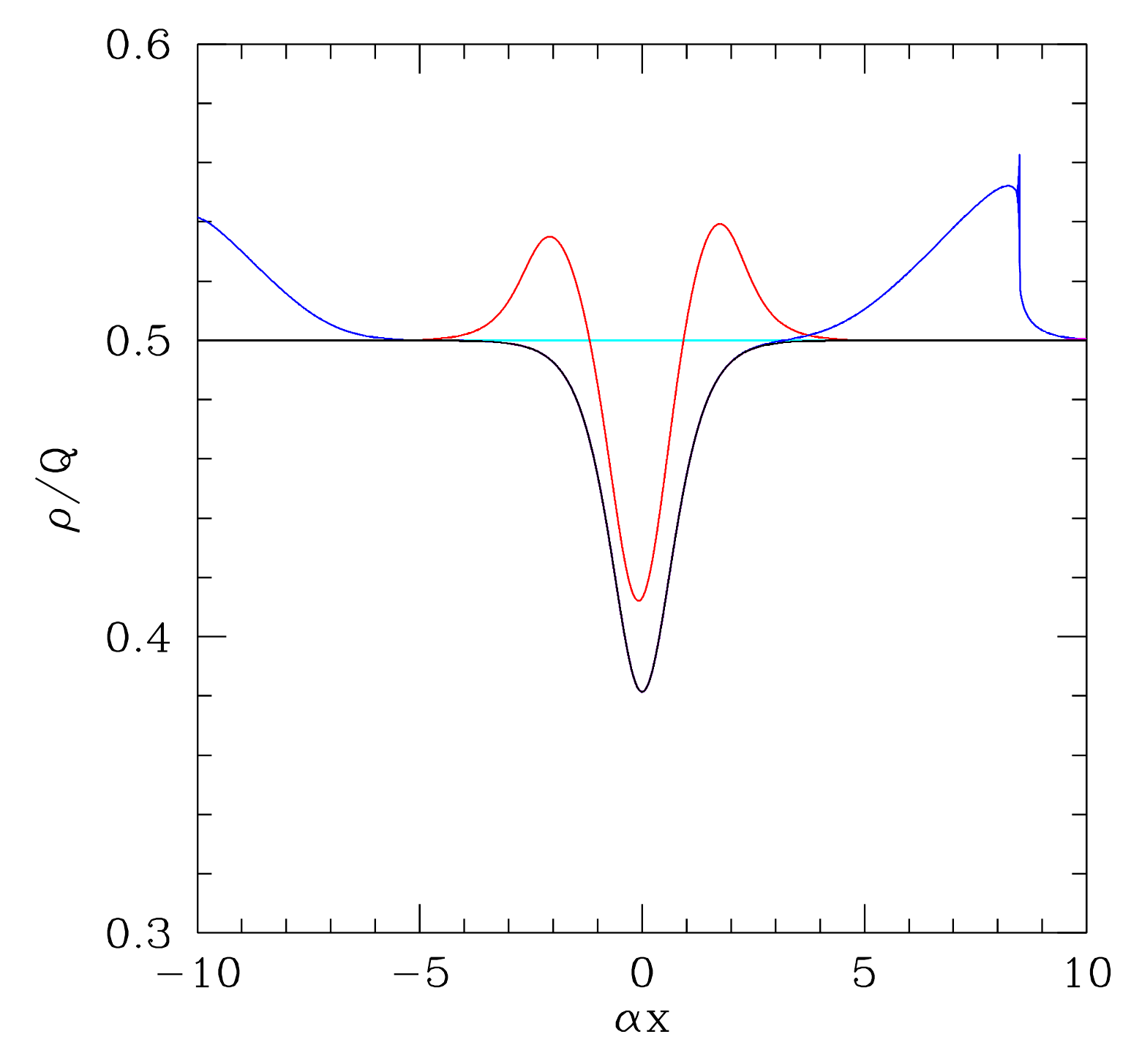}
\caption{Evolution of the density profile for the choice of initial conditions $k_F=\frac{\pi}{2}\,Q$, $k_0=\frac{\pi}{15}\,Q$
obtained by numerical integration of Eqs. (\ref{density}, \ref{phase}). The density profile converges to the analytical, fully subsonic, stationary solution (black line). Colors refer to different times: $\tau=0,1,5,10$ cyan, red, blue, magenta respectively. A localized oscillation is present at $\tau=5$. The curve at $\tau=10$ (magenta) is indistinguishable from the stationary solution (black line) in the range shown in figure.}
\label{fig-gpepi2}
\end{center}
\end{figure}
\begin{figure}[ht!]
\begin{center}
\includegraphics[width=7cm]{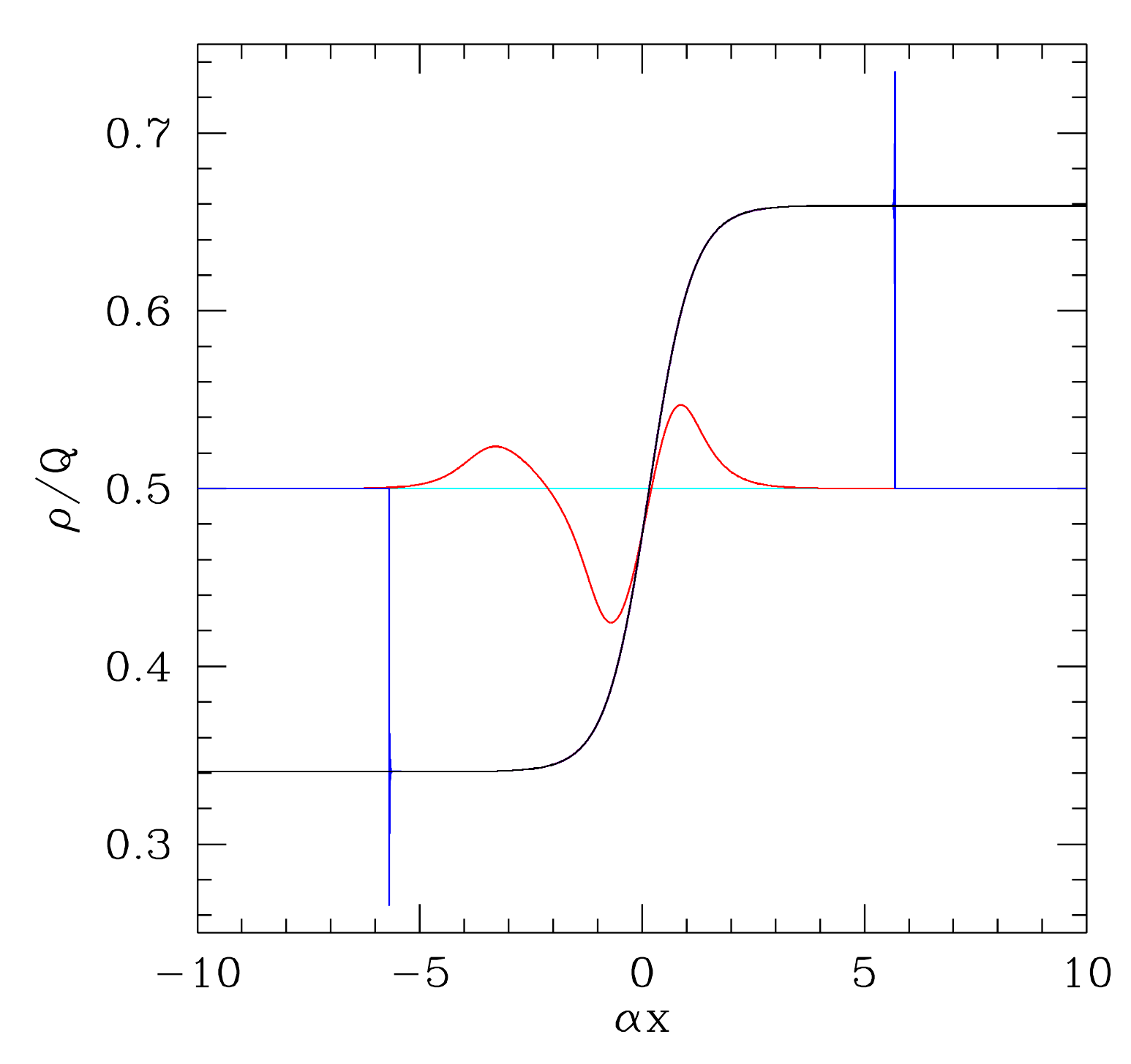}
\caption{Same as Fig. \ref{fig-gpepi2} for the initial conditions $k_F=k_0=\frac{\pi}{2}\,Q$. The density profile converges to the analytical stationary solution (black line). Colors refer to different times: $\tau=0,1,10,20$ cyan, red, blue, magenta respectively. Two shocks (here at $\tau=10$) develop after the quench and travel towards infinity. The curve at $\tau=20$ (magenta) is indistinguishable from the stationary solution (black line) in the range shown in figure.}
\label{fig-gpepi4}
\end{center}
\end{figure}

These equations have been numerically solved verifying the approach to the stationary configuration as shown in Figs. \ref{fig-gpepi2} and \ref{fig-gpepi4} in the subsonic and supersonic case respectively. The semiclassical dynamics shown in Fig. \ref{fig-gpepi4} can be compared with the exact one (for a barrier of width $\alpha=0.1\, Q$) of Fig. \ref{fig-evol-a3}. While in the exact dynamics the density profile is always smooth, in the semiclassical approximation (appropriate for $\alpha \to 0$) two shocks traveling in opposite directions are clearly visible. In both cases, long after the quench, the solution approaches the analytical, stationary state profile. Note that in the semiclassical dynamics the time scale for obtaining a stationary state in a given region of space is roughly a factor two smaller than in the exact solution.

\subsection{The excitation spectrum}
According to the semiclassical approach, phonons are expected to be described by the quantum perturbation around the stationary background configuration, governed by the Hamiltonian $\hat H_2$. The strategy is therefore to diagonalize $\hat H_2$ via a Bogoliubov transformation
\begin{align}
 \delta \hat \psi(x)=\int d\lambda \left\{ D_\lambda (x) \hat a_\lambda +E^*_\lambda (x) \hat a^\dagger_\lambda \right\},
\end{align}
where the annihilation and creation operators $\hat a_\lambda$, $\hat a^\dagger_\lambda$ satisfy the usual commutation relations 
\begin{align}
 [\hat a_\lambda, \hat a^\dagger_\mu]=\delta(\lambda-\mu),
\end{align}
whereas the transformation coefficients must be normalized as
\begin{align}
 \int d\lambda \left\{ D_\lambda(x) D^*_\lambda(x') - E^*_\lambda(x) E_\lambda(x') \right\}&=\delta(x-x'), \cr
 \int d\lambda \left\{ D_\lambda(x) E^*_\lambda(x') - E^*_\lambda(x) D_\lambda(x') \right\}&=0.\label{normalization}
\end{align}
By imposing the diagonalization of $\hat H_2$ in the form
\begin{align}
 \hat H_2=\epsilon +\int d\lambda \,\hbar \omega_\lambda \,\hat a^\dagger_\lambda \hat a_\lambda,
\end{align}
where $\epsilon$ is the fluctuation energy, we are led to the eigenfunction equations
\begin{align}
 h_2 D_\lambda +2g |\psi|^2 \psi^2 E_\lambda &=\hbar \omega_\lambda D_\lambda, \cr
 h_2 E_\lambda +2g |\psi|^2 \psi^{*2} D_\lambda &=-\hbar \omega_\lambda E_\lambda, \label{eigenfunctions}
\end{align}
where we omitted the dependence on $x$ for brevity. Of course, after solving for $D_\lambda$, $E_\lambda$, and $\omega_\lambda$, one must impose the normalization conditions (\ref{normalization}). Notice that if $(\omega, D, E)$ is a solution satisfying the normalization condition, then $(-\omega,E^*,D^*)$ is another formal solution of the same equations, but with negative frequency and negative norm. Moreover, a stable equilibrium solution of (\ref{background}) must allow only for positive eigenfrequencies \cite{pita}.\\
The fluctuation energy is expressed in terms of the solution of the eigenvalue problem as: 
\begin{align}
 \epsilon=-\int d\lambda \, \hbar \omega_\lambda \int dx |E_\lambda(x)|^2.
\end{align}
We can apply this general procedure to two physically relevant cases: the homogeneous state before the quench and the stationary state solution long after the quench. 

Before the quench, in the absence of any external potential, the background configuration is 
\begin{align}
 \psi(x)=\sqrt {\frac {k_F}\pi} e^{-ik_0 x},
\end{align}
corresponding to a uniform density $\rho=\frac {k_F}\pi$ flowing with negative velocity $v=-\frac {\hbar}m k_0$. Substituting this wavefunction in the eigenvalue equations, an exact solution for the coefficients $D_p(x)$ and $E_p(x)$ is given by plane waves. Correspondingly, the fluctuation operator is 
\begin{align}
& 
\delta \hat \psi(x)= \int_{-\infty}^\infty \frac {dp}{N_p} \left[ e^{ipx} \hat a_p +\Gamma_p
e^{-ipx} \hat a^\dagger_p \right] \, e^{-ik_0x}
\label{initial config}\\
& N_p=\sqrt{2\pi (1-\Gamma_p^2)} \\
& \Gamma_p=\sqrt {\xi^2 p^2 +\frac {\xi^4 p^4}4} -\frac {\xi^2 p^2}2 -1.
\end{align}
with healing length $\xi=k_F^{-1}$. The dispersion relation for the excitation spectrum is 
\begin{align}\label{initial spectrum}
 (\omega_p-vp)^2=c^2 \left( p^2+\frac {\xi^2}4 p^4 \right)
\end{align}
where $c=\frac \hbar{m} k_F$. This form coincides with the known Bogoliubov phonon spectrum in a moving fluid $\omega(p)$ \cite{liberati,pita}. 

After the quench the background solution evolves and at later times is defined by the non-homogeneous, asymptotic, stationary solution of the generalized Gross-Pitaevskii equation (\ref{ansatz}, \ref{current}, \ref{amplitude}). In this state, phonons are defined as the normal modes of the fluctuation operator on top of the new background solution. The analysis can be carried out along the same lines followed in the uniform case. In order to solve the eigenfunction equations we make the Ansatz
\begin{align}
\label{ddzero}
 D(x)&=D^0(z) e^{i(\phi(z)+\chi(z))/\alpha}, \\
 E(x)&=E^0(z) e^{-i(\phi(z)-\chi(z))/\alpha},
\end{align}
where we omitted the label $\lambda$, and we set, as usual, $z=\alpha x$. Inserting this form in the linear system (\ref{eigenfunctions}), we obtain an algebraic equation for the spatial derivative of the phase $\chi^\prime(z)$:  
\begin{align}
 (\omega-v\chi')^2=c \left( \chi'^2 +\frac {\xi^2}4 \chi'^4 \right), 
\label{secular}
\end{align}
where
\begin{align}
 v(z)=\frac j{m\rho(z)}, \quad\ c(z)=\pi \frac {\hbar \rho(z)}m, \quad
\xi(z)=\frac \hbar{mc(z)} \nonumber 
\end{align}
are the velocity profile, the sound velocity and the local healing length, respectively. The solution of the secular equation (\ref{secular}) gives two branches
\begin{align}\label{branches}
 \omega_\pm =v\chi' \pm c\sqrt {\chi^{'2}+\frac {\xi^2}4 \chi^{\prime 4}} \, ,
\end{align}
where we omitted the explicit $z$ dependence at right hand side. The corresponding eigenmodes $(D_\lambda^0(z), E_\lambda^0(z))$ are, at fixed $\omega_\lambda$,
\begin{align}
 & E^0_\lambda(z)=\Gamma_\lambda(z)D^0_\lambda(z),\\
 & \Gamma_\lambda=\frac \xi{c} \left( \omega_\lambda -\frac {\xi c \chi^{\prime2}_\lambda}2 -v\chi'_\lambda -\frac c\xi \right),
\label{gamma}
\end{align}
where the label $\lambda$ uniquely identifies each solution of the eigenvalue problem and  $D^0_\lambda$, as determined by the normalization condition (\ref{normalization}), is
\begin{align}
 D^0_\lambda (z)=\left| \frac 1{2\pi} \frac {\partial \chi'_\lambda(z)}{\partial \lambda} \right|^{\frac 12} 
\left[1-\Gamma_\lambda(z)^2 \right]^{-\frac 12}.
\label{dzero}
\end{align}
The normalization condition leads to the condition $\Gamma^2_\lambda(z)<1$, which, due to Eq. (\ref{gamma}), forces the 
choice of the upper sign in Eq. (\ref{branches}). At fixed $\omega >0 $, the analysis of the algebraic equation (\ref{branches}) for $\chi^\prime$ shows that two solutions are always present both in the subsonic and in the supersonic case. However, in the latter case, there are two solutions also for any $\omega >-\omega_{max}(z)$, where
\begin{align}
\omega_{\max}(z)= \frac c{8\xi} \left[ 3\beta-\sqrt {\beta^2+8} \right]\cdot \left[ \beta+4+\sqrt {\beta^2+8} \right]^{\frac 12} \cdot \left[ \beta-4+\sqrt {\beta^2+8} \right]^{\frac 12},
\end{align}
and $\beta(z)=\left \vert \frac{v(z)}{c(z)}\right \vert$ is the local Mach number. This shows that the excitation spectrum and the characteristic frequency $\omega_{max}$ remain finite in the limit $\alpha \to 0$; therefore, the quantity $\alpha \kappa / \omega_{max} \to 0$ (the factor $\alpha$ in the ratio is due to our definition of $\kappa$, which is taken as the derivative of the differential velocity w.r.t. the rescaled length), satisfying the criterion established in Ref. \cite{renaud-omega} for the occurrence of the analogue Hawking radiation. As previously discussed, in the presence of a sonic horizon $\beta(z)$ is monotonic. For negative frequencies $\omega<0$ the mode amplitude diverges at $z^*> 0$ such that $\omega=-\omega_{\max}(z^*)$ and vanishes for $z>z^*$. This discontinuous behavior is likely to be an artifact of the $\alpha\to 0$ limit: we expect that for small but non vanishing $\alpha$ the divergence disappears and the mode decays exponentially for $z>z^*$.\\
The analysis of the Bogoliubov spectrum then shows that the fluctuation operator, long after the quench, is explicitly given in terms of the phonon creation and annihilation operators by
\begin{align}
& \delta\hat \psi(x)=\int \frac {d\lambda}{N_\lambda(z)} 
\left[ \hat b_\lambda +\Gamma_\lambda(z)\, \hat b^\dagger_\lambda \right],\\
& N_\lambda(z)=\sqrt{2\pi\,(1-\Gamma_\lambda(z)^2)}  \, ,
\end{align}
where $\Gamma_\lambda(z)$ is defined in Eq. (\ref{gamma}). Here, $\hat b_\lambda$ and $\hat b^\dagger_\lambda$ are related to the modes $\hat a_\lambda$, $\hat a_\lambda^\dagger$ before the quench by the non-trivial time evolution of the system after switching on the external barrier. 

\subsection{Analogue Hawking radiation}
To relate the phonon operator before and long after the quench we have to consider the evolution from the initial uniform flow, at negative times, to the final stationary state long after the sudden switching on of the smooth barrier (\ref{cosh}). In the Heisenberg picture, the state $|\Phi\rangle$ does not evolve and remains the vacuum state of the Bogoliubov modes $\hat a_p$. The fluctuation operator at any time $t$ can be expressed as a linear combination of the bare modes with time dependent coefficients:
\begin{align}\label{bar expansion}
 \delta\hat \psi(x,t)=\int dp \left\{ D_p(x,t)\,\hat a_p +E^*_p(x,t)\, \hat a ^\dagger_p \right\} \, .
\end{align}
Substituting this expansion into the equation of motion (\ref{eqdeltapsi}) 
we get the equations for the coefficients
\begin{align}
 i\hbar \frac {\partial D_p}{\partial t} &=h_2 D_p+2g|\psi|^2 \psi^2 E_p \, ,\\
 -i\hbar \frac {\partial E_p}{\partial t} &=h_2 E_p+2g|\psi|^{2} \psi^{*2} D_p \, ,
\label{evol-bog}
\end{align}
with initial condition at $t=0$ given by (\ref{initial config}). Looking for a solution that remains regular in the limit $\alpha\to0$, we are led to the Ansatz
\begin{align}
  D(x,t)&=D^0(z,\tau)\, e^{i(\phi(z,\tau)+\chi(z,\tau))/\alpha}, \\
 E(x,t)&=E^0(z,\tau)\, e^{-i(\phi(z,\tau)-\chi(z,\tau))/\alpha},
\end{align}
which indeed satisfies the evolution equations (\ref{evol-bog}) provided that the local momentum $\chi'_p(z,\tau)$ 
is given by
\begin{align}\label{momentum evolution}
 -\frac{\hbar Q}{m}\frac {\partial \chi'_p}{\partial\tau}=\frac{\partial}{\partial z}\,\left [ 
v\chi'_p+c\sqrt{\chi_p^{\prime2}+\frac {\xi^2}4 \chi_p^{\prime4}}\right ] \, ,
\end{align}
where
\begin{align}
 v(z,\tau)&=\frac {\hbar}m \phi'(z,\tau), \\
 c(z,\tau)&=\pi \frac {\hbar}m \rho(z,\tau), \\
 \xi(z,\tau)&=\frac {\hbar}{mc(z,\tau)}
\end{align}
are the time dependent velocity, sound velocity and healing length of the evolving background configuration. $\rho(z,\tau)$ and $\phi'(z,\tau)$ are solutions of the equations (\ref{density}, \ref{phase}), while the initial condition at the quench $\tau=0$ is $\chi^\prime_p(z,0)=p$. 

\begin{figure}
\includegraphics[width=7cm]{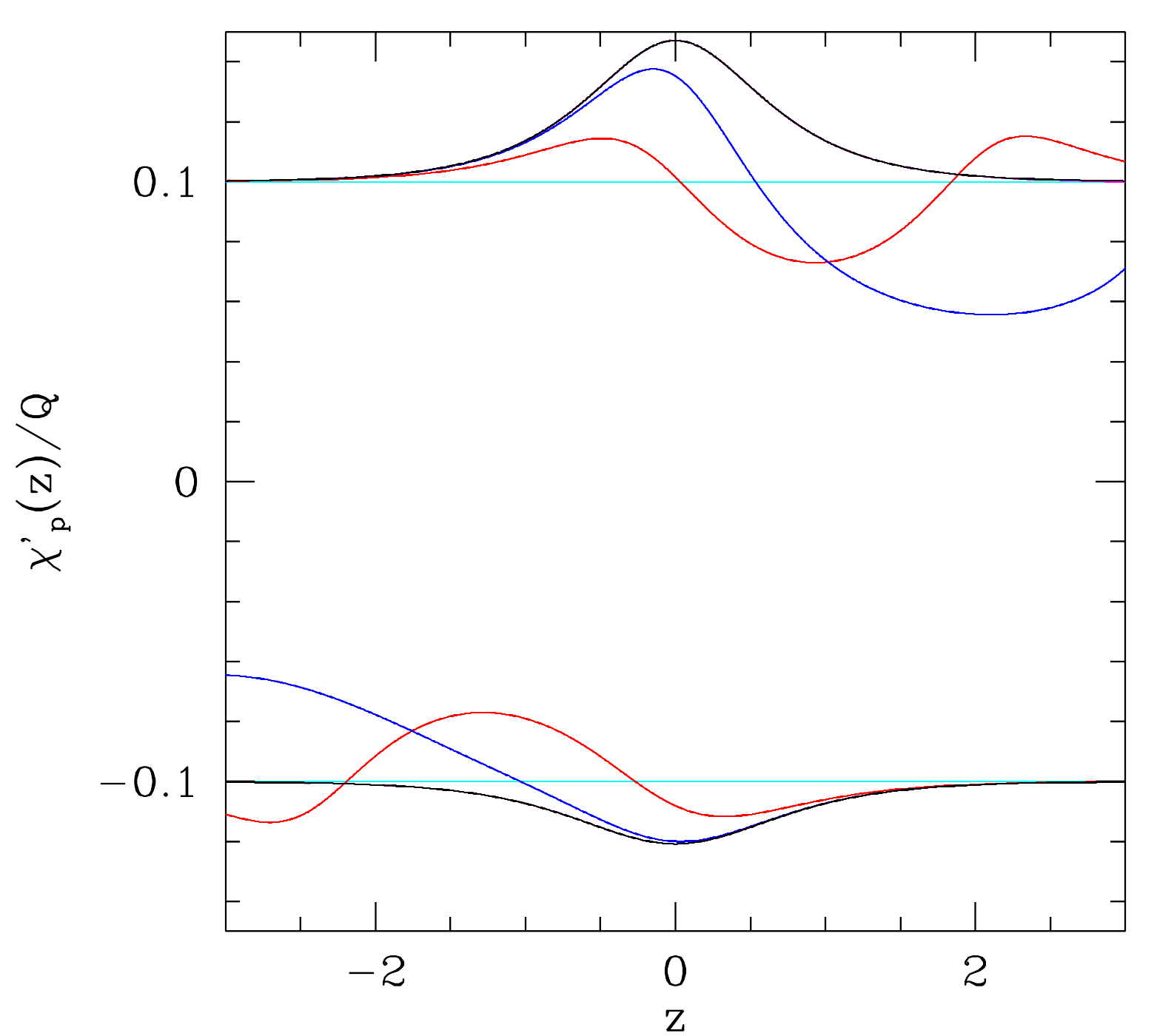}
\caption{Time evolution of the local wavevector after the quench for $k_F=\frac{\pi}{2}\, Q$ and $k_0=\frac{\pi}{15}\,Q$ 
according to Eq. (\ref{momentum evolution}). The upper set of curves correspond to the mode labeled by $p=0.1\,Q$ at the quench. The lower set refers to $p=-0.1 \,Q$. Times are in unit of $\frac{m}{\hbar Q \alpha}$: $\tau=0$ (cyan), $\tau=1$ (red), $\tau=2$ (blue) and $\tau=6$ (magenta). The last line is not visible on the plot because, in the region displayed in the figure, it coincides with the normal mode of the stationary solution (\ref{branches}) corresponding to the same frequency $\omega_p$, shown in black.}
\label{fig-gpex}
\end{figure}

For a parameter choice corresponding to a fully-subsonic, stationary state configuration ($Q < k_F+k_0$) the numerical study of these evolution equations shows that Eq. (\ref{momentum evolution}) has a smooth, regular solution for both positive and negative quasiparticle momenta $p$ approaching, at long times, the expected positive norm solution (\ref{branches}) with the same frequency $\omega_p$ (\ref{initial spectrum}). Two representative examples are shown in Fig. \ref{fig-gpex}. 
In this case we conclude that 
$\hat b_\lambda=\hat a_p$, with $\omega_\lambda=\omega_p$, so the bare operators coincide with the proper normal modes 
also in the final stationary solution.\\
The behavior is qualitatively unchanged also when a sonic horizon forms in the background solution (i.e. for $ k_F-k_0<Q<k_F+k_0$), provided the quasiparticle momentum $p$ is negative. Instead, for positive $p$ the numerical solution of the differential equation (\ref{momentum evolution}) shows the emergence of singularities, implying that the above Ansatz does not correctly describe the actual evolution of the phonon modes. 
\begin{figure}
\includegraphics[width=7cm]{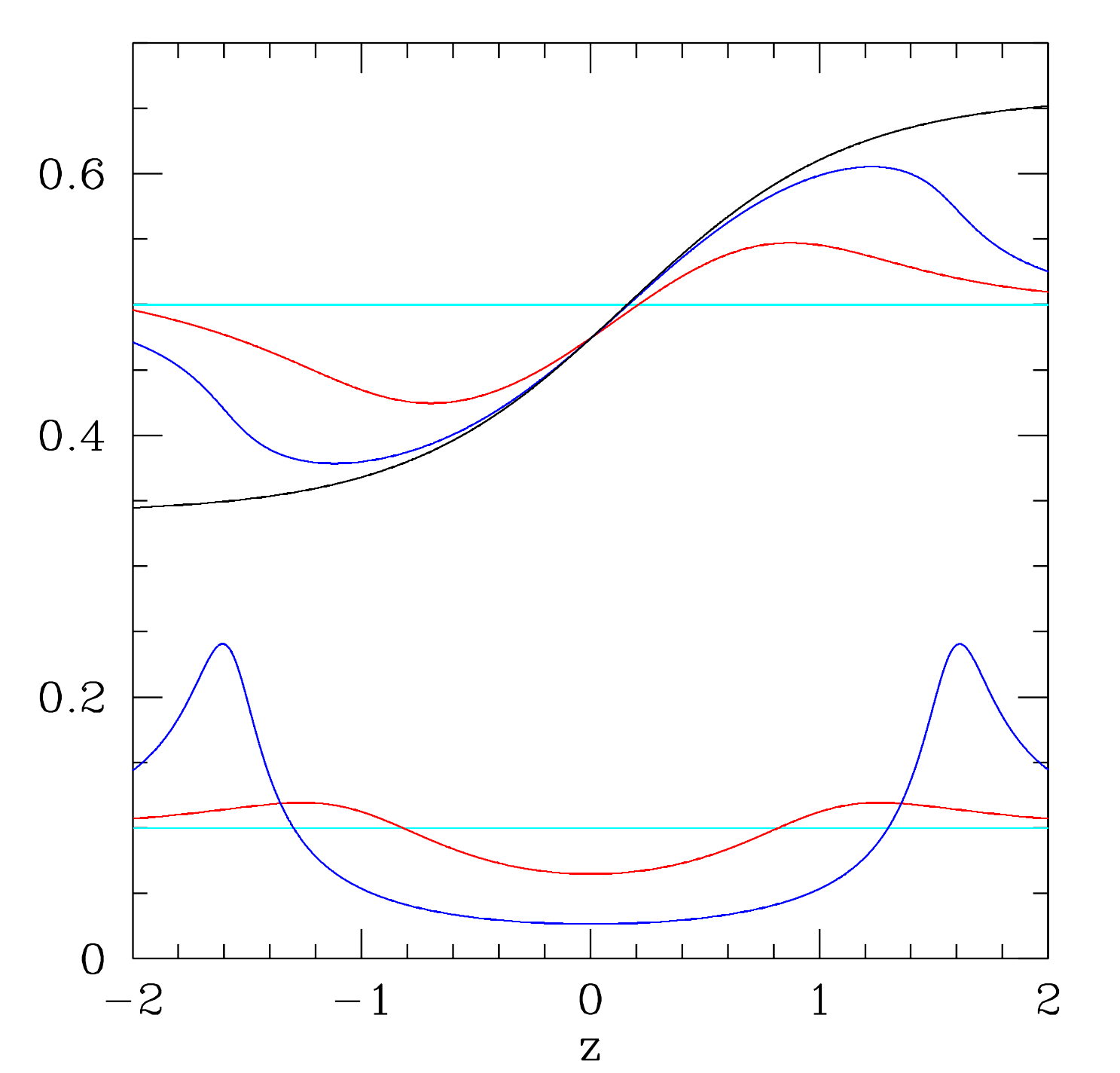}
\caption{Time evolution after the quench for $k_F=k_0=\frac{\pi}{2}\, Q$ according to the semiclassical dynamics
The upper set of curves correspond to the density, the lower set to the local wavevector 
of the mode labelled by $p=0.1\,Q$ at the quench. Both density and wavevector are expressed in units of $Q$. 
Times are in unit of $\frac{m}{\hbar Q \alpha}$: $\tau=0$ (cyan), $\tau=1$ (red), $\tau=2$ (blue). The density profile for the stationary state is shown in black.}
\label{fig-gpeshort}
\end{figure}
A careful numerical study shows that, just after the quench, while the background configuration approaches the stationary solution in a neighborhood of $z=0$, the excitations $D_p(z,\tau)$, $E_p(z,\tau)$ are regular and $\chi'_p(z,\tau)$ preserves the sign of its initial condition $\chi'_p(z,0)=p$, as illustrated in Fig. \ref{fig-gpeshort}. Singularities develop later in time, suggesting that the bare phonon operators $\hat a_p$ differ from the normal modes defined in the stationary background long after the quench. Thus, in this case, we have to identify the unitary transformation relating the initial and the final phonon operators.

To this end we follow the same argument put forward by Hawking in his seminal paper \cite{hawking}: We compare the forward evolution just discussed with the solution of the same equations backward in time, imposing the initial condition at a time $\tau\gg 0$, where the stationary state after the quench has been already reached in a wide portion of space around $z=0$. The fluctuation field operator can then be expressed at all times either in the form (\ref{bar expansion}) or as
\begin{align}
 \delta \hat \psi (x,t)=\int d\lambda \left\{ D_\lambda(x,t) b_\lambda+E^*_\lambda(x,t) \hat b^\dagger_\lambda \right\}.
\end{align}
Matching these two expressions allows to express the bare modes in terms of the phonon operators long after the quench: 
\begin{align}
 \hat a_p=\int d\lambda \left\{ U_{p\lambda} \hat b_\lambda +V_{p\lambda} \hat b^\dagger_\lambda \right\},
\end{align}
with
\begin{align}
 U_{p\lambda}&=\int dx \left\{ D^*_p D_\lambda -E^*_p E_\lambda \right\}, \nonumber \\
 V_{p\lambda}&=\int dx \left\{ D^*_p E^*_\lambda -E^*_p D^*_\lambda \right\}.
\label{unitary}
\end{align}
In these formulas $D$ and $E$ are functions of both $x$ and $t$, but the time dependence must disappear in the final expressions. Therefore, we can choose a convenient time to evaluate the transformation matrices. We fix an optimal time $\tau_0$
when the forward evolution after the quench has led the background function to approximate the asymptotic one in a 
given neighbourhood of $z\sim 0$, let us say $|z|<0.5$ (see Fig. \ref{fig-gpeshort}). \\
To evaluate the unitary transformation (\ref{unitary}) we have to solve the evolution equations (\ref{evol-bog}) both forward and backward in time. It is convenient to consider wave packets of the form 
\begin{align}
\tilde D_{p}(z,\tau) &= \int dp^\prime\, f(p^\prime-p) D_{p^\prime}(z,\tau) \\
\tilde D_{\lambda}(z,\tau) &= \int d\lambda^\prime\, g(\lambda^\prime-\lambda) D_{\lambda^\prime}(z,\tau)
\end{align}
where $f(p)$ and $g(\lambda)$ are weight functions chosen in such a way that, at the initial condition (i.e.  $\tau \to -\infty$ for the forward evolution and $\tau \to +\infty$ for the backward one) the wave packet is centered around $z \sim 0$ (i.e. close to the sonic horizon) with momenta $p$ and $\chi'_{\lambda}(z=0)$, respectively. These wave packets are expected to move away from $z=0$ with a group velocity given by the derivative of the frequency with respect to the wavevector. Now we have to evaluate expressions like
\begin{equation}
\int dz\, \tilde D^*_p(z,\tau_0)\, \tilde D_\lambda(z,\tau_0)
\end{equation}
entering the transformation matrices. Note that in order to give a significant contribution to the integral, both wave packets, at time $\tau_0$, must be localized the same region. However, during the time evolution from the initial condition up to the matching time $\tau_0$, the wave packet $\tilde D_\lambda$, that moves backward in time, will proceed to the right because waves cannot propagate upstream after the sonic horizon, at least at long wavelengths. Analogously, $\tilde D^*_p$, moving forward in time, will be dragged inside the black hole. The only possibility to find both $\tilde D_\lambda$ and $\tilde D^*_p$ centered in the same region is to require that both group velocities are vanishingly small, i.e., that the momenta $p$ and $\chi'_{\lambda}(z=0)$ are positive and small. For this choice of parameters we can assume that $D^0_\lambda(z,\tau_0)$ is still well-approximated by its value long after the quench, i.e., by the expression (\ref{dzero}) appropriate for the normal mode corresponding to the stationary state solution (\ref{ansatz}, \ref{current}, \ref{amplitude}). For $\omega_\lambda\sim 0$, $z\sim 0$ and $\chi'_\lambda(z)>0$ we get
\begin{align}
 \chi'_\lambda(z)&\sim  \frac {\omega_\lambda}{\kappa z} \qquad (\omega_\lambda z>0),\\
 D^0_\lambda(z) &\sim \frac 1{\sqrt{4\pi \xi(0) |\omega_\lambda|}}, \\
 \Gamma_\lambda(z) &\sim -1+\xi(0) \,\chi'_\lambda(z),
\end{align}
where
\begin{align}
 \kappa=\left. \frac {d}{dz} [c(z)+v(z)]\right|_{z=0}
\end{align}
is the Unruh analogue surface gravity (recall that in our model $v(z)$ is negative). The divergence of $\chi'_\lambda(z)$ when $z\to0$ implies that the leading contribution to the integrals defining $U_{p\lambda}$ and $V_{p\lambda}$ comes from such a region, where we can approximate 
$D^0_p(z,\tau_0)\sim D^0_p(0,\tau_0)$:
\begin{align}
 U_{p\lambda}&=C_{p\lambda}\int_{-\infty}^\infty  dz\, 
e^{\frac i\alpha \left( \frac {\omega_\lambda}\kappa \log |z|-\chi'_p(0)z \right)} ,\\
 V_{p\lambda}&=-C_{p\lambda}\int_{-\infty}^\infty  dz \,
e^{ -\frac i\alpha \left( \frac {\omega_\lambda}\kappa \log |z|+\chi'_p(0)z \right)} ,\\
 C_{p\lambda}&=\frac {D^0_p(0,\tau_0)}{\alpha \sqrt{4\pi \xi(0)|\omega_\lambda|}},
\end{align}
which, after integration, give the transformation matrices
\begin{align}
 U_{p\lambda}&= -iG^+_{p\lambda} \left| \frac {\chi'_p(0)}{\alpha} \right|^{-\left(1+i\frac {\omega_\lambda}{\alpha\kappa}\right)} e^{\frac {\pi|\omega_\lambda|}{2\alpha\kappa}},\\
 V_{p\lambda}&= iG^-_{p\lambda} \left| \frac {\chi'_p(0)}{\alpha} \right|^{-\left(1-i\frac {\omega_\lambda}{\alpha\kappa}\right)} e^{-\frac {\pi|\omega_\lambda|}{2\alpha\kappa}},
\end{align}
with
\begin{align}
 G^\pm_{p\lambda}=({\rm sign}\ \omega_\lambda)\, C_{p\lambda} \,\Gamma\left( 1\pm i\frac {\omega_\lambda}{\alpha\kappa} \right). 
\end{align}
Recall that, since we are working in the Heisenberg picture, the ground state $|\Phi\rangle$ is unchanged during the time evolution. However, it is non-trivially related to the vacuum $|0\rangle$ of the phonon operators long after the quench, defined
by $\hat b_\lambda |0\rangle=0$: 
\begin{align}
 |\Phi \rangle =e^{-\frac 12 \int d\lambda d\lambda' F_{\lambda\lambda'} \hat b^\dagger_\lambda \hat b^\dagger_{\lambda'}} |0\rangle,
\end{align}
where the symmetric matrix $F_{\lambda\lambda'}$ is defined as the solution of the linear problem
\begin{align}
 \int d\lambda \, U_{p\lambda} \, F_{\lambda\lambda'}=V_{p\lambda'}.
\end{align}
By substituting the previous expressions we finally get 
\begin{align}
 F_{\omega\omega'}=\delta(\omega+\omega') e^{-\pi\frac {|\omega|}{\alpha\kappa}},
\end{align}
and we see that it mixes positive and negative frequencies. The ground state, in terms of the dressed phonon operators, has thus the form of a two-mode squeezed state. Notice that the negative frequencies are only defined in the supersonic region and cannot reach the region $z>0$. By tracing out them \cite{yurke,sciama} we then get a thermal density matrix with the Hawking temperature
\begin{align}
 k_B T_H=\alpha\frac {\hbar \kappa }{2\pi},
\label{thawk}
\end{align}
which is the same obtained in the exact solution of the model (\ref{th}).\\
Let us notice that $|\Phi\rangle$ is written as a linear superposition of excited states including an infinite number of phonons of positive and negative frequencies. Since each pair has zero energy, this combination remains an eigenstate of the fluctuation Hamiltonian $\hat H_2$, which is essential to prove that the asymptotic state is a stationary solution of the evolution equations.

\section{Experimental configurations}
\begin{figure}
\begin{center}
\includegraphics[width=7cm]{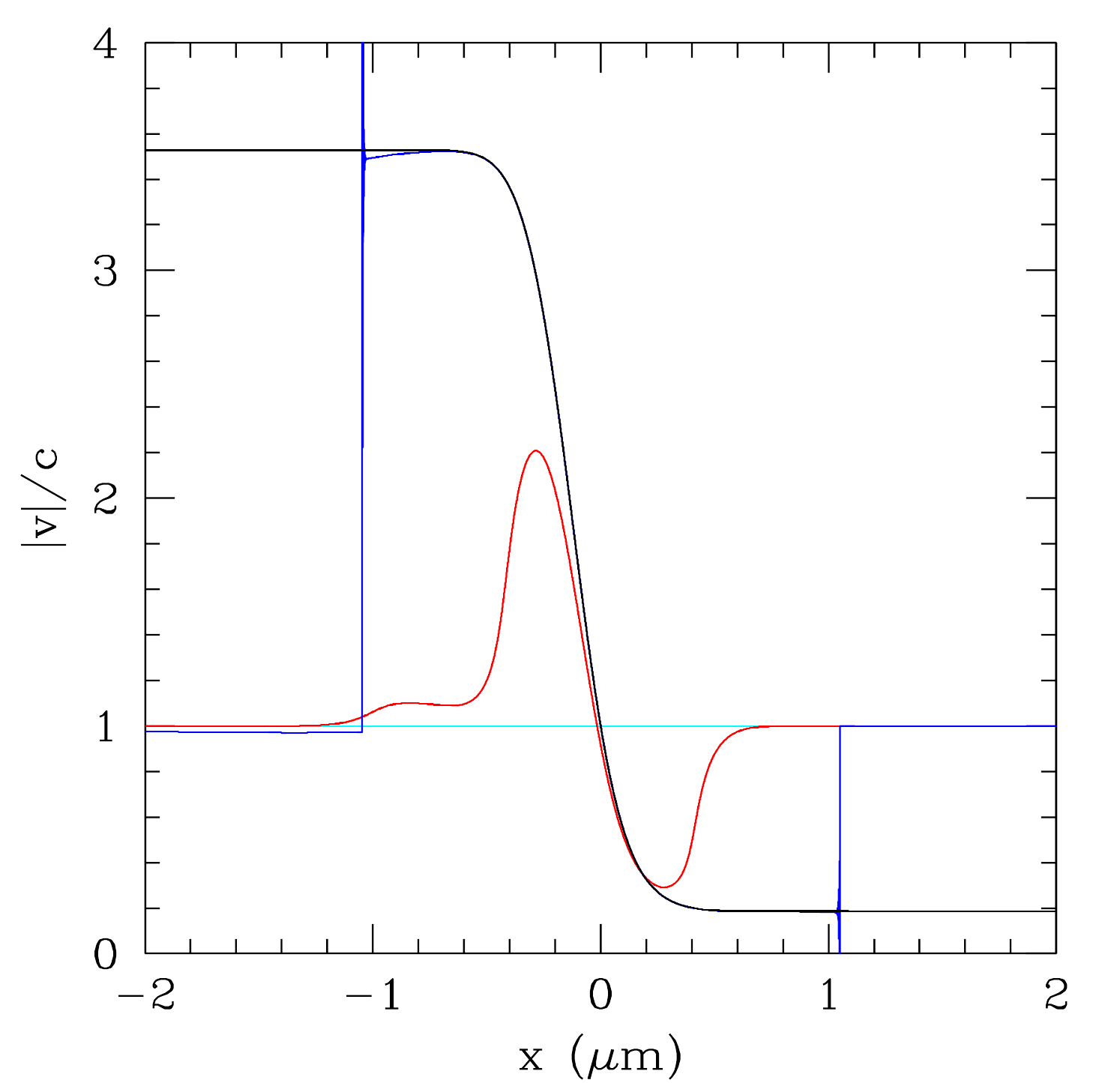}
\caption{Semiclassical time evolution of the Mach ratio after the quench for the parameters discussed in the text. 
Different colors correspond to different times after the quench: $t=0$ (cyan), $\tau=0.02$ ms (red), $\tau=0.08$ ms (blue), $t=0.2$ ms (black). The curves gradually approach the stationary state profile in the whole range shown in the figure.}
\label{fig-gpe3}
\end{center}
\end{figure}

The exhaustive analysis we have presented refers to the specific case of Hard Core Bosons in one dimension, because only in this limit an analytic solution is available. Although this special system has been already experimentally reproduced several years ago \cite{kinoshita,bloch}, the Tonks-Girardeau limit is rather difficult to obtain due to the tight one-dimensional confinement required. We believe that the strong coupling restriction may be relaxed and the Hawking emission may be seen also under less stringent conditions, as recent studies suggest. To this aim, we briefly examine the case of a gas of $^{87}$Rb atoms confined in a cylindrical trap of radius $a_\perp =0.25\,\mu$m, as an illustrative example. The trap is assumed to be flat in the longitudinal direction with total length $L\gtrsim 10 \mu$m. Let us fix the initial number density of the condensate to $\rho_0=3.8\,10^3\,\mu$m$^{-1}$ and its initial velocity to $v_0\sim 18$ mm/s. Now we turn on an external potential of the form $V(x)=V_0\,e^{-(\alpha\,x)^2}$ with $V_0 = \frac{\hbar^2Q^2}{2m}$ and $\alpha =0.1\,Q$. The potential amplitude corresponding to $Q=38\,\mu$m$^{-1}$ is $V_0\sim 3.6\,\mu$K. Then, we follow the semiclassical dynamics by integrating the usual (cubic) Gross-Pitaevskii equation in the $\alpha\to 0$ limit. As illustrated in Fig. \ref{fig-gpe3} a sonic horizon forms in $x=0$ and the flow reaches a stationary state in a region of $10\,\mu$m around the horizon after a fraction of a millisecond. 
According to Eq. (\ref{thawk}) we thus expect the emission of Hawking radiation at the relatively high temperature $T_H\sim 100$ nK. Measuring the momentum distribution in the far upstream region or the density correlations across the horizon, the characteristic  signatures of thermal emission should be clearly visible. In an experimental set-up it might be more convenient to maintain the condensate still and move the external barrier, as in the experiments of Ref. \cite{steinhauer1}. This choice is clearly equivalent, via Galileo transformation, to the case considered in the previous Sections.

\section{Conclusions}

The exact solution of a one-dimensional model of interacting bosons allows to unveil the microscopic origin of the phonon emission representing the condensed matter analogue of the Hawking radiation. Without referring to the gravitational analogy, we investigate the phonon flux which originates in the stationary state reached by a flowing BEC in the presence of an external potential. The fully analytical solution of the model provides several advantages over previous studies in this field: in fact, we unambiguously show that thermal emission can be expected only for a condensate flowing against a  barrier potential which accelerates the particles to supersonic speeds and only in the limit of extremely smooth obstacle, i.e., when the semiclassical (WKB) limit is reached. Otherwise phonon emission does occur (even in the absence of a sonic horizon) but the thermal character is lost. This is the case of an external potential which has a step-like form. Moreover, we are able to follow the quench dynamics mimicking the experimental procedure, confirming that the analytical stationary solution is indeed reached at long times. Since density correlations across the horizon represent a widely adopted probe for the occurrence of the Hawking mechanism, we evaluate the correlation pattern; this shows the features expected on the basis of semiclassical arguments. 

Although the exact solution contains all the required information about the model, we deem instructive to also develop a semiclassical study of the same set-up, in order to understand whether the currently adopted semiclassical paradigm, based on the use of the Gross-Pitaevskii equation and of the Bogoliubov analysis of the excitation spectrum, is able to reproduce the exact results. Surprisingly we find that, even in the presence of extremely smooth potentials, the semiclassical analysis lacks some important features: indeed, an entire class of stationary solutions, characterized by non-monotonic velocity profiles, cannot be obtained by means of semiclassical arguments.

Finally, we provide an illustrative example of a possible experimental set-up apt to show thermal phonon emission in a BEC at a reasonably high temperature. 

The formalism introduced here is suited to further generalizations in order to address other phenomena often studied in the field, such as the black-hole laser effect or the dynamical Casimir effect. To this extent, innovative results could be developed in these directions also. Unfortunately, the extension of this model to higher dimensions is not trivial.

\appendix
\section{}\label{app:spectrum}
Here we report the spectrum of the non-interacting single-particle Hamiltonian
\begin{equation}
h=-\frac{\hbar^2}{2m}\frac{d^2}{dx^2} +V(x) \, ,
\label{hamf}
\end{equation}
with $V(x)=\frac{\hbar^2Q^2}{2m}\Theta(x)$. The eigenvalues are written in the free particle form $\epsilon_k = \frac{\hbar^2k^2}{2m}$, while the corresponding wave-functions acquire different forms for $0<k \le Q$ or $|k|>Q$:

$i)$ $0<k \le Q$. In this regime the spectrum is non-degenerate and the eigenfunctions read:
\begin{equation}
\phi_{k}(x) =\begin{cases} 
\frac{1}{\sqrt{2\pi}}\,\left [ e^{ikx}+R_k\,e^{-ikx}\right ] & \mbox{for } x < 0 \\
\frac{1}{\sqrt{2\pi}}\,T_k\,e^{-\lambda_kx}& \mbox{for } x > 0 
\end{cases} ,
\nonumber 
\end{equation}
where $\lambda_k=\sqrt{Q^2-k^2}$ and the reflection and transmission coefficients are
\begin{equation}
R_k = \frac{k-i\lambda_k}{k+i\lambda_k} \, , \qquad T_k = \frac{2k}{k+i\lambda_k} \, .
\end{equation}
These states are exponentially trapped in the region $x<0$. Note that the eigenfunctions $\phi_k(x)$ for $-Q<k \le 0$ are not defined.

$ii)$ $|k|>Q$. Here the spectrum is doubly degenerate. The right moving wave for $k>Q$ is:
\begin{equation}
\phi_{k}(x) =\begin{cases} 
\frac{1}{\sqrt{2\pi}}\,\left [ e^{ikx}+R_k\,e^{-ikx}\right ] & \mbox{for } x < 0 \\
\frac{1}{\sqrt{2\pi}}\,T_k\,e^{iqx}& \mbox{for } x > 0 
\end{cases}
\nonumber 
\end{equation}
and the left moving wave for $k<-Q$ is: 
\begin{equation}
\phi_{k}(x) =\begin{cases} 
\sqrt{\frac{|k|}{2\pi q}}\,T_k\,e^{ikx} & \mbox{for } x < 0 \\
\sqrt{\frac{|k|}{2\pi q}}\,\left [ e^{-iqx}+R_k\,e^{iqx}\right ] 
& \mbox{for } x > 0 
\end{cases},
\nonumber
\end{equation}
where $q=\sqrt{k^2-Q^2}$ and the reflection and transmission coefficients are:
\begin{equation}
R_k = \frac{k-q}{k+q} \, \qquad T_k = \frac{2k}{k+q} \, ,
\end{equation}
for the right moving and
\begin{equation}
R_k = -\frac{k+q}{k-q} \, , \qquad T_k = -\frac{2q}{k-q} \, .
\end{equation}
for the left moving.
The normalization condition between any pair of wavefunctions reads, as usual,
\begin{equation}
\int_{-\infty}^\infty dx\,\phi_k(x)^* \phi_{k^\prime}(x) = \delta(k-k^\prime) \, .
\end{equation}

\section{}\label{app:longtime}

We give some detail on the evaluation of the asymptotic form (\ref{asy}) starting from the exact time evolution (\ref{psit}). We first evaluate the internal products $\langle \phi_p| \psi^0_k \rangle$ (\ref{internal}) which, using the explicit expressions of Appendix A, are written as a sum of simple poles like
\begin{equation}
\frac{\alpha_p}{\eta \pm i(p-k)} = \alpha_p \pi\delta(p-k) \mp P\frac{i\alpha_p}{p-k} \, ,
\nonumber 
\end{equation} 
where the Plemelj identity has been used. Inserting this sum into Eq. (\ref{psit}) we get:
\begin{equation}
\pi \alpha_k \phi_k(x)\,e^{-\frac{i}{\hbar}\epsilon_k t}  \mp i\, 
P\,\int_{-\infty}^{\infty} dp \,\frac{\alpha_p\phi_p(x)e^{-\frac{i}{\hbar}\epsilon_p t}}{p-k},
\nonumber 
\end{equation}
where, besides the term coming from the $\delta$-function, the integral gives also a non-vanishing contribution at long times. In fact, by defining $u=p-k$ and expanding $\alpha_{k+u}$ and $\epsilon_{k+u}$ to first order in $u$, it is straightforward to show that the second term has a finite limit as $t\to +\infty$: 
\begin{equation}
\mp\pi\alpha_k \phi_k(x)\,e^{-\frac{i}{\hbar}\epsilon_k t} \,{\rm sgn} (\epsilon_k^\prime) \, ,
\nonumber 
\end{equation}
whose sign depends on the slope of the fermionic dispersion $\epsilon_k$ so that, for $\pm\epsilon_k^\prime > 0$, it precisely cancels the contribution coming from the $\delta$-function. In conclusion, we can formally write:
\begin{equation}
\lim_{t\to +\infty} \frac{e^{-\frac{i}{\hbar}\epsilon_p t}}{\eta\pm i(p-k)} = 
2\pi\delta(k-p)\,e^{-\frac{i}{\hbar}\epsilon_k t}\, \Theta(\mp\epsilon_k^\prime).
\nonumber
\end{equation}
This analysis is easily extended to all external potentials which do not admit bound states, because the pole contribution in the internal product (\ref{internal}) comes from the asymptotic regions $x\to\pm\infty$, where every eigenfunction $\phi_p(x)$ acquires the form of a scattering state, i.e., can be written as a linear combination of an incident (or transmitted) and a reflected wave. Therefore, the structure of the eigenfunctions at $x\to\pm\infty$ has the same form shown in Appendix A, irrespective of the details of the external potential. 

\section{}\label{app:exact}
Exact results for the barrier (\ref{cosh}) can be obtained starting from (\ref{hyper}) and using that
\begin{align}
 |\Gamma(1+ik)|^2 &= \frac {k\pi}{\sinh {k\pi}}, \\
 |\Gamma(\textstyle{\frac 12} + ik)|^2 &= \frac {\pi}{\cosh {k\pi}},
\end{align}
as well as the asymptotic expression
\begin{align}
 F(a,b;c;z)\sim \frac {\Gamma(c)\Gamma(c-a-b)}{\Gamma(c-a)\Gamma(c-b)} +\frac {\Gamma(c)\Gamma(a+b-c)}{\Gamma(a)\Gamma(b)} (1-z)^{c-a-b}
\end{align}
when $z\to 1$ (since in our case Re$(c-a-b)=0$). 
These allow to find exact expressions for the asymptotic density $\rho(x)$ when $x\to\pm\infty$:
\begin{align}
\rho_+=\frac {k_F}\pi+\frac \alpha{4\pi^2} \left[ \frac {1+\coth \frac {2\pi Q}\alpha}{\sinh \frac {2\pi Q}\alpha} \log \frac {\frac \pi\alpha (Q+k_F+k_0)}{\frac \pi\alpha (Q-k_F-k_0)} \right.+ \left. \coth \frac {\pi Q}\alpha \log \frac {\frac \pi\alpha (Q-k_F+k_0)}{\frac \pi\alpha (Q+k_F-k_0)} \right], \\
 \rho_-=\frac {k_F}\pi+\frac \alpha{4\pi^2} \left[ \frac {1+\coth \frac {2\pi Q}\alpha}{\sinh \frac {2\pi Q}\alpha} \log \frac {\frac \pi\alpha (Q+k_F-k_0)}{\frac \pi\alpha (Q-k_F+k_0)} \right.+\left. \coth \frac {\pi Q}\alpha \log \frac {\frac \pi\alpha (Q-k_F-k_0)}{\frac \pi\alpha (Q+k_F+k_0)} \right].
\end{align}

\section{}\label{app:hyper}
According to the exact form (\ref{hyper}) of the scattering states for a barrier of the form (\ref{cosh}), the hypergeometric function $F(a,b;c;\zeta)$ has to be evaluated for arguments $(a,b,c)$ whose imaginary part grows large as $\alpha\to 0$. Here we derive the appropriate asymptotic limit of the hypergeometric function. First we use the identity (see Ref. \cite{abramo}, Eq. 15.3.3):
\begin{equation}
F(a,b;c;\zeta) = (1-\zeta)^{c-a-b} F(c-a,c-b;c;\zeta) .
\nonumber
\end{equation}
Then we note that, in the case of interest (\ref{hyper}), the variable $z$ is real and the parameters are related by $2c-a-b=1$. Therfore, we can express the hypergeometric function in terms of Legendre functions by (see Ref. \cite{abramo} Eq. 15.4.17):
\begin{equation}
F(a,b;c;\zeta) = \Gamma(c)\,\left [\zeta(1-\zeta)\right]^{\frac{1-c}{2}} P^{1-c}_{a-c}(1-2\zeta).
\nonumber
\end{equation}
Finally, recalling that $\zeta = \frac{1-\tanh(\alpha x)}{2}$ belongs to the interval $(0,1)$, we use the integral representation (see Ref. \cite{grad} Eq. 8.714-2) to obtain:
\begin{eqnarray}
F(a,b;c;\zeta) = \frac{\Gamma(a+b)}{\Gamma(a)\Gamma(b)}\,
 \int_0^\infty dt\,\frac{t^{a-1}}{\left [1+2(1-2\zeta)t+t^2\right ]^{c-\frac{1}{2}}}.
\nonumber 
\end{eqnarray}
Inserting the expressions for $a=\frac{1}{2} -i \frac{k+Q}{\alpha}$ and 
$b=\frac{1}{2}-i\frac{k-Q}{\alpha}$, the integral can be written as
\begin{equation}
\int_0^\infty dt \frac{e^{i\frac{\varphi(t)}{\alpha}}}
{\sqrt{t\left [1+2(1-2\zeta)t+t^2\right ]}},
\end{equation} 
with 
\begin{eqnarray}
\varphi(t)= -(k+Q)\, \log t + k\,\log \left [1+2(1-2\zeta)t+t^2\right ].
\end{eqnarray}
These expressions hold for any value of $\alpha$. However, in the $\alpha\to 0$ limit they considerably simplify because the integral can be explicitly evaluated by use of the stationary phase method. Looking for the extrema of $\varphi(t)$ with $t>0$, we have to consider two distinct regimes:
\begin{itemize}
\item If $k > Q$, the unique extremum is given by 
\begin{equation}
t_- = \frac{-Q\,\tanh(\alpha x) - \Delta}{Q-k}.
\end{equation}
\item Instead, if $k < Q$, two solutions exist for $Q\,\tanh(\alpha x) < -\sqrt{Q^2-k^2}$:
\begin{equation}
t_\pm = \frac{-Q\,\tanh(\alpha x) \pm \Delta}{Q-k},
\end{equation}
while no extremum is present for  $Q\,\tanh(\alpha x) > -\sqrt{Q^2-k^2}$ ,
\end{itemize}
where, in both cases,  
\begin{equation}
\Delta = \sqrt{Q^2 (\tanh(\alpha x))^2 +k^2-Q^2}.
\end{equation}
The second derivative of the phase, evaluated at the extremum, is given by
\begin{equation}
\varphi^{\prime\prime}(t_\pm) = \frac{k^2-Q^2}{kt_\pm^2}\frac{\Delta}{\Delta \mp k\,\tanh(\alpha x)}.
\end{equation}
Therefore, each extremum (if any) contributes to the integral with a term
\begin{equation}
\sqrt{\frac{\alpha\pi}{\Delta}}\,e^{i\frac{\varphi(t_\pm)}{\alpha}}\,e^{\mp i\frac{\pi}{4}}.
\end{equation}
The scattering states (\ref{hyper}) then acquire an analytic form in the $\alpha\to 0$ limit. By use of the Stirling formula for the asymptotic behavior of the $\Gamma$ function, we finally obtain: 
\begin{equation}
\phi_k(x) = \sqrt{\frac{k}{2\pi\Delta}}\, 
\left [ \zeta\left (1-\zeta\right )\right ]^{-i\frac{k}{2\alpha}} 
\left \{ e^{i\frac{\varphi_-}{\alpha}} -i e^{i\frac{\varphi_+}{\alpha}}\right \},
\nonumber 
\label{asinto0}
\end{equation}
where we used the shorthand notation $\varphi(t_\pm)=\varphi_\pm$. The second term in curly brackets is present only for $k < Q$ and $Q\,\tanh(\alpha x) < -\sqrt{Q^2-k^2}$, while for $k < Q$ and $Q\,\tanh(\alpha x) > -\sqrt{Q^2-k^2}$ the wavefuncion vanishes to leading order in $\alpha$. This result has been derived for $k>0$. The analogous expression for $k < 0$ is simply obtained by changing $k\to |k|$ and $x\to -x$ (or $\zeta \to 1-\zeta$). 

As $\alpha\to 0$, the square modulus of the scattering wavefunction just reduces to 
\begin{equation}
{\hskip -0.3cm} |\phi_k(x)|^2 = \frac{1}{2\pi} \frac{k}{\sqrt{Q^2 [\tanh(\alpha x)]^2 +k^2-Q^2}}
\label{asinto1}
\end{equation}
for $k >Q$, while for $k<Q$ and  $Q\,\tanh(\alpha x) < -\sqrt{Q^2-k^2}$ an oscillatory term still survives:
\begin{eqnarray}
|\phi_k(x)|^2 = \frac{1}{\pi} \frac{k}{\sqrt{Q^2 [\tanh(\alpha x)]^2 +k^2-Q^2}} \left [ 1 + \sin\frac{\varphi_+-\varphi_-}{\alpha}\right ] .
\label{asinto2}
\end{eqnarray}
These rapid oscillations in the single-particle scattering wavefunction are however washed out when we perform th integration over the wavevectors $k$ required to evaluate the averages in the Fermi gas (\ref{1b}).



\begin{thebibliography}{99}
\bibitem{note} Note that also the bosonic momentum density (or current) operator coincides with the fermionic one because it can be written in terms of the commutator between the density and the Hamiltonian.
\bibitem{nota-step} Usually the definition of ``waterfall potential'' refers to a smooth step instead of a sharp one. We will add this modification later on, showing that our conclusions hold in this case also.
\bibitem{error} Due to a misprint, the second term in this expression was disregarded in Ref. \cite{epl}.
\end{thebibliography}
\end{document}